\documentclass[conference,compsoc]{IEEEtran}
\IEEEoverridecommandlockouts

\usepackage{url} 
\usepackage{framed}
\usepackage{mdframed}
\usepackage{listings}
\usepackage{multicol}
\usepackage{amssymb}
\usepackage{lipsum}
\usepackage{adjustbox}
\usepackage[font=bf,skip=\baselineskip]{caption}
\usepackage{tabularx}
\usepackage{seqsplit}
\usepackage{multirow}
\usepackage{booktabs}
\usepackage{graphicx}
\usepackage{hyperref}
\usepackage{subcaption}
\usepackage{makecell}
\usepackage{xcolor}
\usepackage[ruled,vlined,linesnumbered] {algorithm2e}
\usepackage{amsmath}
\usepackage{tikz}
\usepackage{tcolorbox}
\usepackage{threeparttable}
\usepackage{tablefootnote}
\usepackage{csquotes}
\usepackage{colortbl}
\usepackage{algorithm2e}
\usepackage{adjustbox}
\usepackage{balance}
\usepackage{dblfloatfix}
\usepackage{placeins}
\usepackage{afterpage}

\usepackage{breakurl} 
\usepackage{boldline}
\usepackage[ruled,vlined]{algorithm2e}
 
\usepackage{pifont}
\usepackage{balance}

\usepackage{enumitem}
\usepackage[htt]{hyphenat}

\definecolor{customcite}{HTML}{FF0000}
\definecolor{customlink}{HTML}{00BB00}
\definecolor{customurl}{HTML}{0066CC}
\AtEndPreamble{
\usepackage{hyperref}

    \hypersetup{
      colorlinks = true,
      linkcolor = customlink,
      anchorcolor = purple,
      citecolor = customcite,
      filecolor = purple,
      urlcolor = customurl
    }
}

\definecolor{customcell}{HTML}{ffebb7}

\newcommand{\yes}{\tikz\draw[fill=black] (0,0) circle (.35em);}
\newcommand{\no}{\tikz\draw[fill=white] (0,0) circle (.35em);}
\newcommand{\partialcell}{\begin{tikzpicture}
    \filldraw[fill=white] (0,0) circle (.35em);
    \filldraw[fill=black] (0,.35em) arc (90:270:.35em);
\end{tikzpicture}}

\newcommand{\serv}{\textit{LSF}}

\begin{document}

\title{Demystifying Real-world LLM Service Freeloading}

\author{Anonymous Author(s)}

\author{
\IEEEauthorblockN{Xinyi Hou, Shenao Wang, Wenhui Pan, Yanjie Zhao, and Haoyu Wang}
\IEEEauthorblockA{
Huazhong University of Science and Technology, Wuhan, China\\
xinyihou@hust.edu.cn, shenaowang@hust.edu.cn, wenhuipan@hust.edu.cn,\\ yanjie\_zhao@hust.edu.cn, haoyuwang@hust.edu.cn}
}

\maketitle

\begin{abstract}
Third-party projects repackage free inference from LLM web applications and dedicated clients into independent APIs. We refer to such use beyond providers' permitted access, allocation, or redistribution conditions as LLM service freeloading (LSF). However, its scale and underlying mechanisms remain poorly understood.
We present a systematic measurement study of 513 open-source projects, tracing 1,588 inference paths across 64 upstream LLM services and characterizing their development, dependencies, and operating mechanisms. We find that 93.0\% of repositories were created in 2026. Direct HTTP or WebSocket requests account for 83.2\% of paths, making reproduction of product protocols the dominant invocation method. Projects adapt messages, reconstruct conversation state, and convert responses to support compatible APIs. Account selection or rotation appears on 304 paths (19.1\%), allowing projects with multiple eligible accounts to distribute requests across their separate allowances. Our policy analysis finds that 58 of the 62 services with available terms explicitly prohibit at least one examined practice, including automated access, account sharing, or service redistribution. These restrictions cover both downstream API provision and the means of obtaining upstream inference.
Our findings inform defenses against unauthorized consumption and redistribution of free LLM resources. We have responsibly disclosed our findings to the relevant providers.
\end{abstract}

\section{Introduction}
\label{sec:introduction}

Large language models (LLMs) have become an integral part of various AI applications, supporting tasks such as question answering, content generation, and software development~\cite{chatterji2025chatgpt,hou2024llm4se}. To make these capabilities accessible to users, many AI providers integrate model inference into their applications through web interfaces or dedicated clients. AI chat assistants (e.g., ChatGPT and DeepSeek) offer web-based access to model inference~\cite{deepseekwebsite,chatgpt}, and AI coding assistants (e.g., Claude Code and Codex) provide inference capabilities through dedicated development tools and clients~\cite{claudecodeoverview,codexauth}. Alongside paid offerings, many LLM applications provide free inference access through free-tier allowances, limited trials, or promotional grants. Such access remains subject to eligibility requirements, usage limits, and restrictions on automation, sharing, and redistribution~\cite{chatgptpricing,openaiterms,claudecodelegal}. These conditions determine who may consume the free inference resources, through which access channels, and for which purposes.

Unfortunately, recent security reports have uncovered how inference access provided for specific applications can be repurposed to serve unrelated workloads~\cite{vercel2026tokentheft}. Third-party implementations can expose this access through independently operated APIs, allowing downstream clients to consume inference through interfaces outside the original product. We refer to the repackaging of free inference access from LLM product channels into independent APIs in violation of upstream access or usage conditions as \textbf{LLM Service Freeloading~(\serv{})}.
For example, Vercel~\cite{vercel2026tokentheft} reported \textit{Chipotlai Max}~\cite{chipotlai2026}, which converts Chipotle's customer-support chatbot into a backend for coding clients.  Okta~\cite{okta2026tokens} further analyzed \textit{Poison Claude}, which advertises discounted inference by aggregating AWS promotional free bonus credits. Such practices can divert free inference resources to unintended workloads, extend access beyond eligible users or permitted limits, and obscure which downstream clients consume those resources. However, to our knowledge, little has been done so far to systematically explore \serv{} in the wild and understand the ecosystem behind it.

To address this gap, we present a systematic measurement study of LLM service freeloading in the wild. Starting from a manually curated seed set, we expand discovery through large-scale automated repository searches and source-code analysis, constructing a dataset of 513 open-source projects with 1,588 inference paths across 64 upstream LLM services. We characterize their development and upstream dependencies, analyze how they repackage free inference, and examine the provider conditions governing these practices. Overall, we have the following findings.

First, \textbf{\serv{} is growing rapidly, with concentrated popularity, development activity, and upstream dependencies.} Of the 513 repositories, 93.0\% were created in 2026. Leading projects attracted substantial attention, with the most-starred repository receiving 75,841 stars and the ten most-starred projects accounting for 89.0\% of all stars. Development was similarly concentrated, with the ten projects contributing the most post-creation commits accounting for 41.5\% of such activity. Code reuse is also very common, with 1,828 shared commits identified across 90 projects. We identified 64 upstream LLM services whose free inference access was repackaged by \serv{} projects. DeepSeek Web, Gemini Web, Qwen Web, and ChatGPT Web were common sources, with 264 projects (51.5\%) repackaging access to at least one of these four services.

Second, \textbf{provider restrictions extend beyond redistribution to the mechanisms supplying inference.} Among the 62 upstream services with available terms, 58 explicitly prohibited at least one practice associated with \serv{}. Counting both prohibitions and conditional restrictions, automated access was restricted by 57 services (91.9\%), redistribution by 50 (80.6\%), limit circumvention by 49 (79.0\%), account sharing by 48 (77.4\%), and multiple-account use or repeated acquisition of free allowances by 35 (56.5\%). These findings show that the permissibility of a downstream API depends on how it obtains and sustains upstream inference, as well as whether redistribution itself is permitted.

Third, \textbf{\serv{} commonly relies on direct product requests, protocol adaptation, and account management.} Direct HTTP or WebSocket requests accounted for 1,322 of the 1,588 inference paths (83.2\%), making reproduction of product protocols the dominant means of accessing upstream inference. Browser-mediated access and dedicated-client bridges provided alternative invocation methods. To connect these product channels to downstream APIs, 794 paths (50.0\%) reconstructed prompts or conversation histories, while 210 emulated tool calls by converting generated text into structured responses. Projects also reused and coordinated upstream access through credential import and persistence, observed on 298 paths, and account selection or rotation, observed on 304 paths (19.1\%). These mechanisms allow implementations to retain product authentication state and, when multiple eligible accounts are configured, distribute requests across their separate allowances. 

\noindent\textbf{Contributions.}
We make the following contributions:
\begin{itemize}[leftmargin=15pt]
    \item \textbf{Measurement dataset.}
    We construct a dataset of 513 open-source LSF projects, comprising 1,588 inference paths across 64 upstream LLM services, providing a basis to understand this ecosystem. 

    \item \textbf{Ecosystem and policy analysis.}
    We characterize project growth, code reuse, development activity, popularity, and upstream dependencies, revealing how attention and maintenance concentrate within the ecosystem. We further examine five categories of provider restrictions to assess the conditions governing the acquisition, consumption, and redistribution of free inference.

    \item \textbf{Mechanism analysis.}
    We organize LSF implementations into seven functional groups and analyze how credential management, product workflow reconstruction, protocol adaptation, and account scheduling support independent inference APIs. Through detailed case studies, we explain how these mechanisms operate together and identify their implications.
\end{itemize}

\section{Background}
\label{sec:background}

\subsection{LLM Services and Access Policies}
\label{subsec:llm-service}

\noindent\textbf{LLM services.}
LLM services make model inference available to users and software applications without requiring them to operate the underlying models~\cite{chatterji2025chatgpt,sergeyuk2025coding}. Model providers expose inference through developer APIs~\cite{openaiapi}, while cloud and inference-hosting platforms, such as Amazon Bedrock\cite{bedrockoverview} and Hugging Face\cite{hfinferenceservices}, provide managed access to models from different developers. Multi-model platforms, such as OpenRouter~\cite{openrouteroverview}, further bring multiple upstream services behind a common interface and route requests across inference providers.
At the application level, these capabilities support conversational assistants such as ChatGPT~\cite{chatgpt}, search assistants such as Perplexity~\cite{perplexityoverview}, and coding agents~\cite{sergeyuk2025coding,zhang2026coding} such as Codex~\cite{codex} and Claude Code~\cite{claudecodeoverview}. Inference is also embedded in customer-support systems~\cite{aditya2026customeragent}, office software~\cite{wang2024officeagent}, and content-creation tools~\cite{notionaiwriting}, where it supports specific functions within a broader product~\cite{intercomfinfaq,microsoftcopilotoverview}. Users interact with these offerings through web and mobile applications, desktop clients, development tools, or model playgrounds, while the services manage the underlying inference requests~\cite{claudecodeoverview,hfinferenceservices,microsoftcopilotoverview}. 

\noindent\textbf{LLM service policies.}
LLM providers establish policies governing the allocation, billing, and permitted use of inference resources. Access may be offered through free allowances, subscription plans, or usage-based billing, with promotional programs granting credits to eligible users~\cite{chatgptpricing,codexauth,okta2026tokens}. Our study focuses on free inference provided through product-level access channels, including free tiers, limited trials, and promotional grants. The conditions attached to these offerings determine who qualifies for access and whether benefits may be claimed repeatedly. Usage constraints further govern request rates, cumulative consumption, and access to particular models~\cite{chatgptpricing,codexauth}. Separate conditions regulate permitted access methods and whether accounts or inference capabilities may be shared, integrated into other applications, or redistributed to third parties~\cite{openaiterms,claudecodelegal}. These policies therefore govern both the amount of free inference available and the ways in which it may be obtained and provided to downstream clients.

\begin{figure}[t]
    \centering
    \includegraphics[width=1\linewidth]{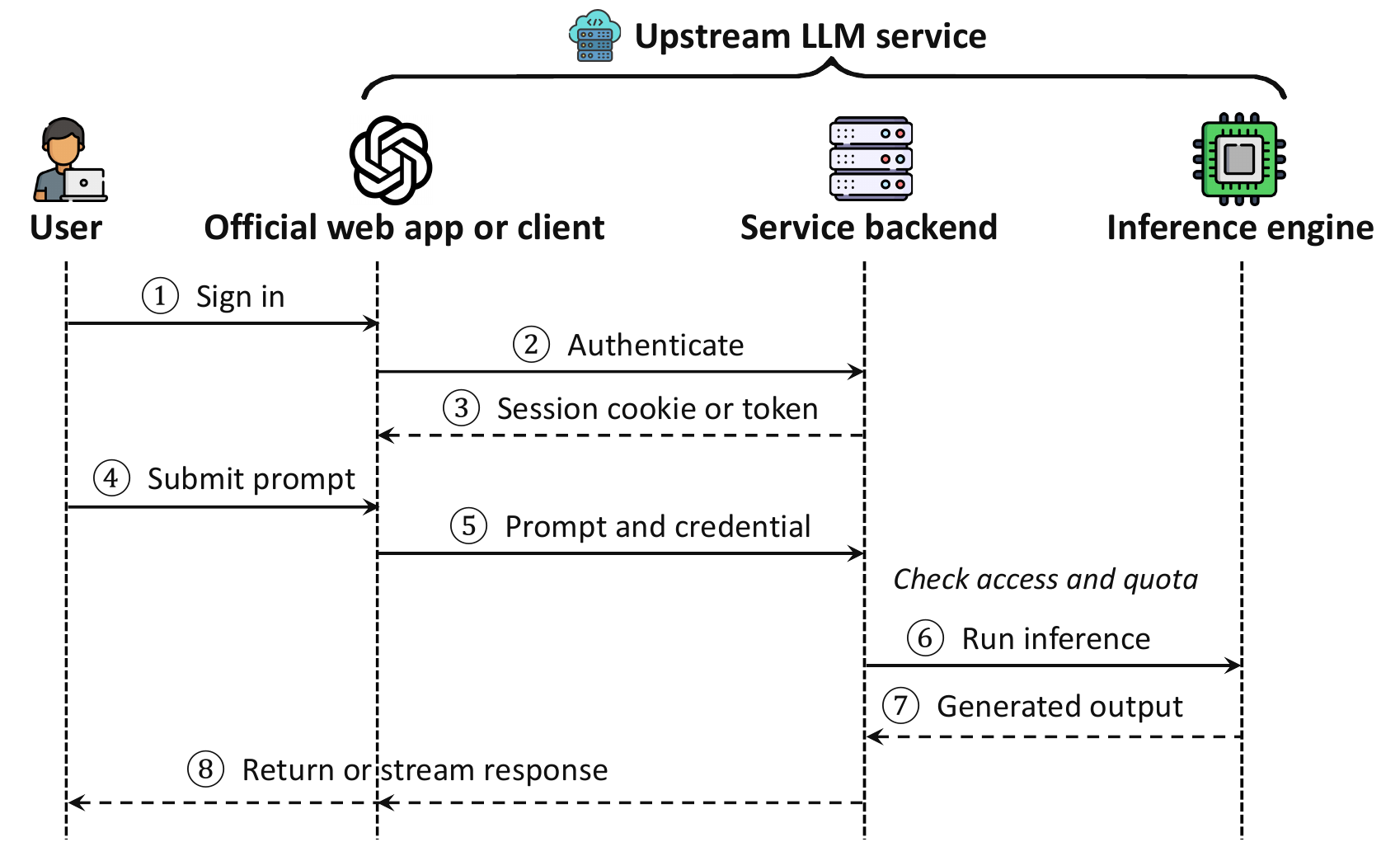}
    \caption{A typical authenticated inference request through an LLM service's web application or dedicated client. The backend checks the user's credentials and available quota before invoking the model and returning the result through the client.}
    \label{fig:normal_workflow}
\end{figure}

\noindent\textbf{A typical access workflow.}
\autoref{fig:normal_workflow} illustrates a typical authenticated workflow for accessing inference through an LLM application. A user first signs in through the application's web interface or dedicated client. Upon successful authentication, the service establishes a session or issues an access token~\cite{rfc6265,rfc6749,codexauth,claudecodeauth}. When the user submits a prompt or task, the client constructs a request to the application's inference endpoint and includes the corresponding session cookie or access token. The backend validates these credentials, identifies the associated account, and checks the applicable permissions and usage limits. An accepted request is then passed to the inference service, and the generated output is returned through the client, either as a complete response or incrementally through streaming. Resource consumption is recorded against the associated account under its applicable quota or billing arrangement. For developer API access, callers submit requests using API credentials associated with their account or project, which similarly determine the permissions and resource allocation applied to each request~\cite{openaiapi}.

\subsection{Abuse of Free LLM Inference}
\label{subsec:free-inference-abuse}

\begin{figure}[t]
    \centering
    \includegraphics[width=1\linewidth]{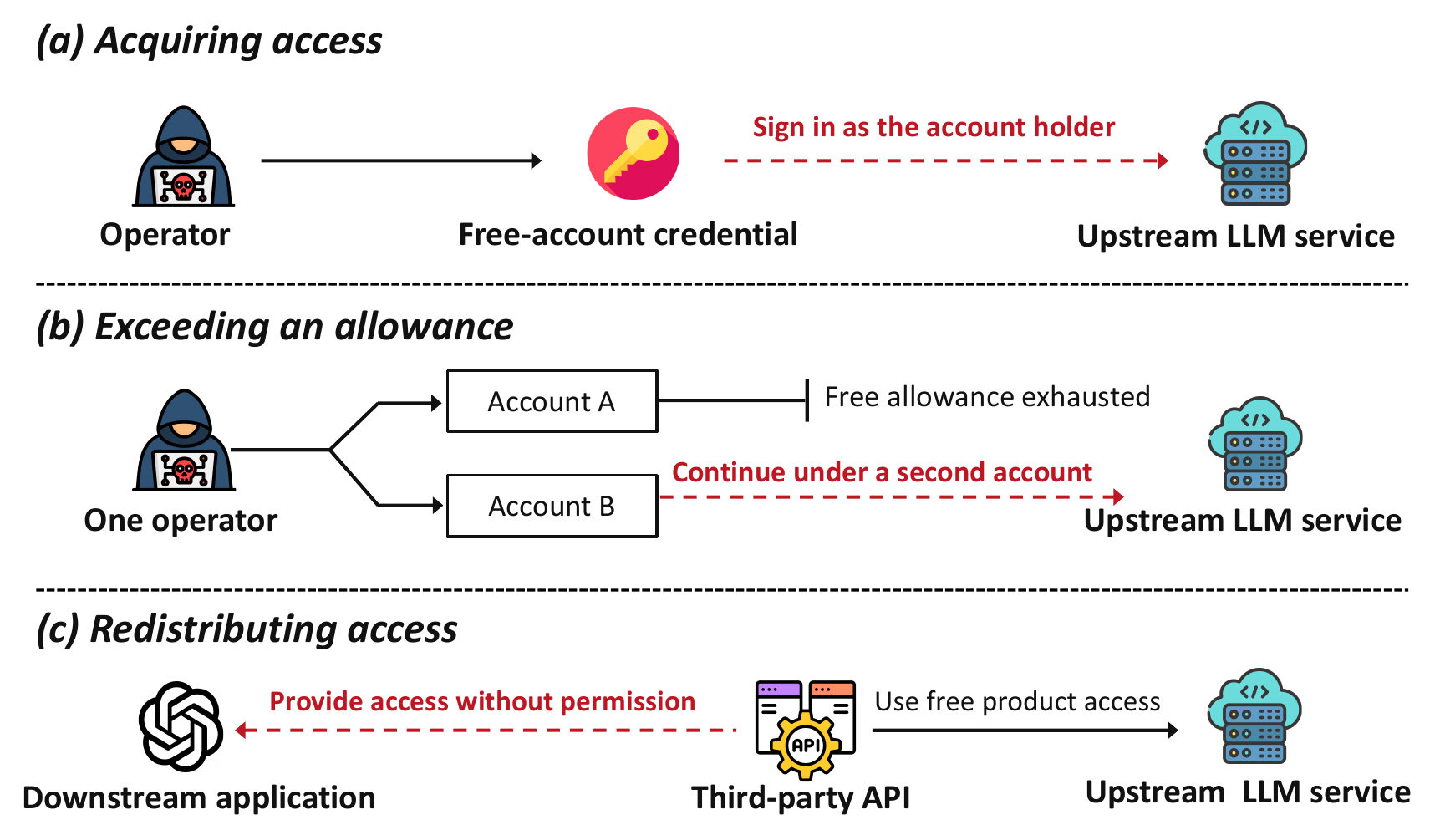}
    \caption{Examples of abuse involving free LLM inference. \textcolor{red}{Red} dashed arrows indicate (a) using stolen credentials to access another user's free inference resources, (b) switching accounts to exceed a per-user free allowance, and (c) exposing free product access through an independent API without authorization.}
    \label{fig:llm_service_abuse}
\end{figure}

Many LLM providers offer free inference through free tiers, limited trials, or promotional grants. These resources can be abused when actors circumvent the conditions governing their provision. \autoref{fig:llm_service_abuse} illustrates examples involving the acquisition of free access, consumption beyond permitted allowances, and unauthorized redistribution.

\noindent\textbf{Acquiring access and entitlements.}
Actors may obtain free inference access by circumventing registration and eligibility requirements or accessing other users' accounts without authorization. Enrollment abuse can involve creating multiple accounts to repeatedly claim benefits restricted to one user, or using invalid payment methods to obtain trial access~\cite{okta2026tokens,stripe_fraud_2026}. These practices allow an operator to acquire free allowances beyond those permitted under the offering's eligibility and enrollment conditions.

\noindent\textbf{Consuming resources beyond their permitted scope.}
Once free access is available, abuse can affect both the workloads it supports and the amount of inference consumed. Application-specific access may be redirected toward unrelated tasks, such as using customer-support inference for coding workloads~\cite{vercel2026tokentheft}. Usage controls may also be circumvented or modified to obtain inference beyond the assigned allowances~\cite{sourcegraph2023abuse}. \autoref{fig:llm_service_abuse}(b) illustrates an operator switching accounts to continue consuming inference after exhausting a per-user free allowance. The individual requests may pass authentication while the combined consumption exceeds the limit applicable to that user.

\noindent\textbf{Redistributing access.}
Free product access can be exposed through \textit{LLM relay services}, which accept downstream API requests and forward them to upstream inference channels. A relay may retain upstream credentials while independently managing downstream authentication, usage limits, and charges~\cite{teamcymru2026relay}. Interface adaptation and request routing allow downstream clients to use the upstream product's free inference through a separate API. Redistribution becomes abusive when this arrangement violates the conditions governing who may use the free access or how it may be provided to others. 

\subsection{Definition and Scope of \serv{}}
\label{subsec:definition}

In this study, \serv{} denotes the repackaging of free inference access intended for an LLM service's web application or dedicated client into an independent third-party API. We refer to software projects that implement this functionality as \serv{} projects.
Formally, let $E$ denote a third-party service implementation, $P$ an upstream LLM service, $\mathcal{C}_P$ its set of product-level access channels, and $a_E$ a downstream API endpoint. We classify $E$ as a \serv{} implementation if there exist $P$, $c \in \mathcal{C}_P$, and $a_E$ such that
\begin{equation}
\label{eq:caire}
\begin{aligned}
\serv{}(E) \iff {}&
\mathrm{FreeAccess}(E,P,c)\\
&\land \mathrm{Expose}(E,a_E)\\
&\land \mathrm{Repackage}(E,P,c,a_E).
\end{aligned}
\end{equation}

G1--G3 specify the technical requirements. All predicates must hold along the same inference path under a supported configuration.

\noindent\textbf{G1: Upstream LLM service.}
$\mathrm{FreeAccess}(E,P,c)$ requires that $E$ obtains free inference resources from $P$ through a product-level channel $c$. These resources include free-tier allowances, trial access, promotional grants, and anonymous free access. The upstream service may also offer paid plans.

\noindent\textbf{G2: Downstream LLM API.}
$\mathrm{Expose}(E,a_E)$ requires that $E$ provides an independent API through which downstream clients submit inference requests and receive generated responses. The API may use a standard or custom protocol and may serve local or remote clients.

\noindent\textbf{G3: Inference repackaging.}
$\mathrm{Repackage}(E,P,c,a_E)$ requires that requests received through $a_E$ trigger inference at $P$ through $c$, and that the corresponding results return through the downstream API. This includes both direct forwarding and implementations that translate requests, maintain service state, or adapt responses.

\noindent\textbf{Scope of problem.}
We exclude inference paths relying solely on paid subscription allowances, developer APIs, or locally hosted models. Authorized repackaging and the use of free resources within their permitted conditions do not constitute \serv{}. Misuse involving only prohibited generated content, as well as attacks intended solely to exhaust resources or impose financial harm, is also outside our scope.

\subsection{Threat Model}
\label{subsec:threat-model}

We consider an operator of a \serv{} implementation as defined in \autoref{subsec:definition}.

\noindent\textbf{Protected resources.}
The protected resources are the inference capacity and free-use entitlements allocated by upstream services. The security objective is to preserve their use by eligible users and workloads under the applicable access and resource-allocation conditions.

\noindent\textbf{Adversary objective.}
The adversary seeks to obtain and sustain free upstream inference for requests submitted through an independently operated API, beyond the permissions or allowances granted by the upstream service. The resulting inference may support the operator's own applications or be provided to other clients.

\noindent\textbf{Adversary capabilities.}
The adversary controls the downstream API implementation and the components that interact with upstream services. They can construct requests, manage upstream sessions, and select the accounts or access channels used to serve each request. Upstream access may rely on legitimate free accounts, anonymous sessions, or access obtained without authorization. These capabilities do not require control over the upstream models or inference infrastructure.
\section{Dataset Construction}
\label{sec:methodology}

\begin{figure}[htbp]
    \centering
    \includegraphics[width=1\linewidth]{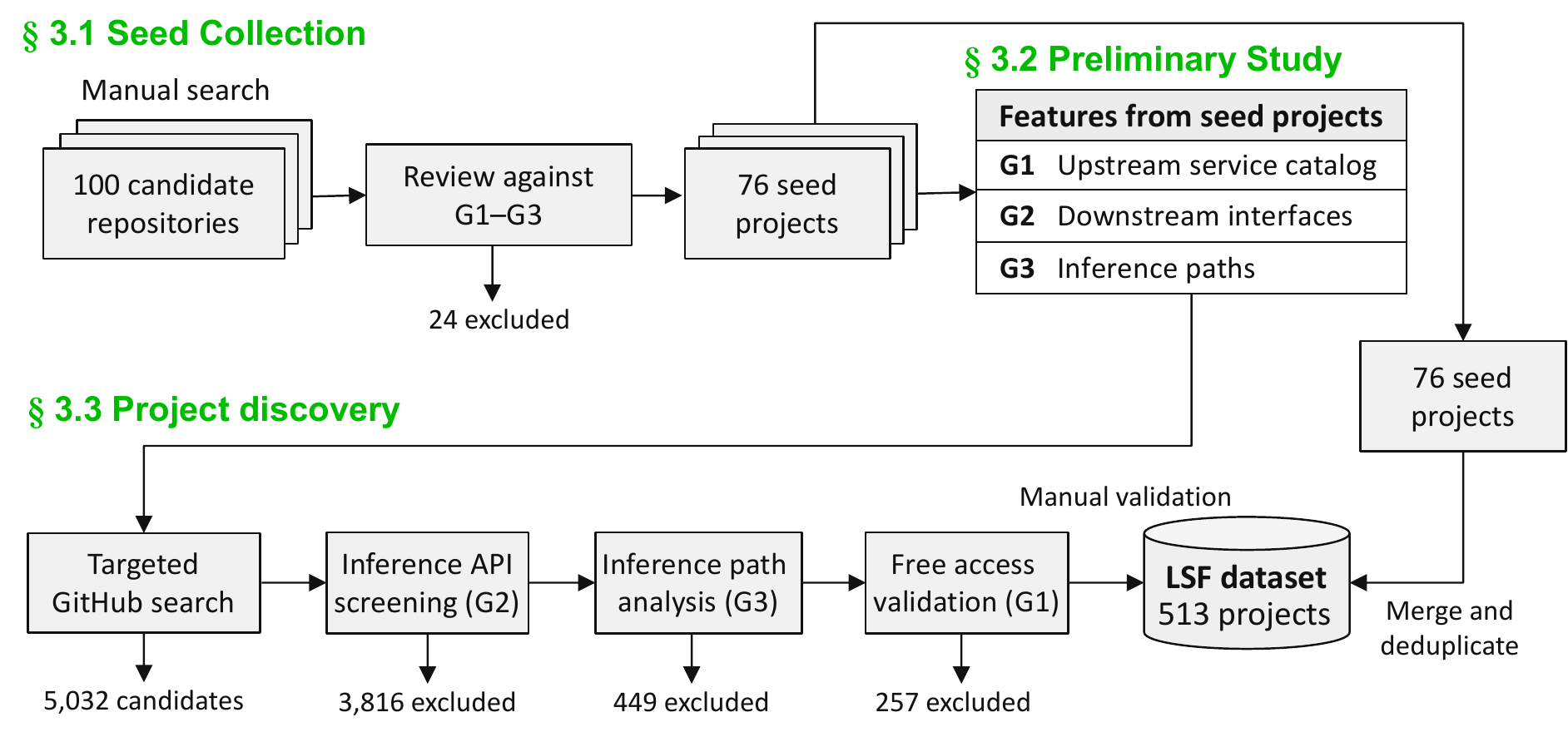}
    \caption{Construction of the \serv{} dataset.}
    \label{fig:dataset_construction}
\end{figure}

In this section, we construct a dataset of open-source \serv{} projects from a manually annotated seed set and an expanded search of GitHub repositories and source code, as shown in \autoref{fig:dataset_construction}. We include a repository if at least one supported inference path satisfies G1--G3 in \autoref{subsec:definition}.

\subsection{Seed Collection}

We constructed an initial candidate set through manual searches on platforms including GitHub, Google, Reddit, and LINUX DO. Queries included phrases such as ``free llm api'' and ``web to llm api'', combined with LLM service names. On GitHub, we examined repository descriptions and READMEs for indications of free inference access. Searches on Google, Reddit, and LINUX DO helped locate additional candidates referenced in public project lists and technical discussions. This search yielded an initial collection of 100 GitHub repositories for manual examination.
Two researchers with experience in LLM application security and program analysis jointly examined 30 repositories to develop annotation guidelines based on G1--G3 in \autoref{subsec:definition}. For each repository, we assessed whether the implementation (1) obtains free inference resources through an upstream LLM service's web application or dedicated client, (2) exposes an independent inference API to downstream clients, and (3) connects that API to the identified upstream channel and returns the corresponding inference results. We inspected project documentation, configuration examples, and relevant source code, consulting upstream service documentation to establish the availability of free access. The researchers recorded the supporting documentation and code locations and discussed ambiguous cases to establish consistent inclusion and exclusion rules.

Using these guidelines, the two researchers independently annotated the remaining 70 repositories, achieving a Cohen's $\kappa$ of 0.919 on project inclusion decisions before adjudication. A third researcher with five years of research experience in LLM security reviewed disagreements and helped resolve them through discussion. The annotation identified 76 \serv{} projects and excluded 24 repositories that did not satisfy the definition. Examples outside our scope include resource lists (\href{https://github.com/mnfst/awesome-free-llm-apis}{\texttt{mnfst/awesome-free-llm-apis}}), gateways for free-tier developer APIs (\href{https://github.com/DevvGwardo/free-llm-router}{\texttt{DevvGwardo/free-llm-router}}), and chat applications using local models (\href{https://github.com/vemonet/libre-chat}{\texttt{vemonet/libre-chat}}). These 76 \serv{} projects formed our seed set for deriving the textual and code patterns used in subsequent project discovery.

\begin{figure}[t]
    \centering
    \includegraphics[width=0.99\linewidth]{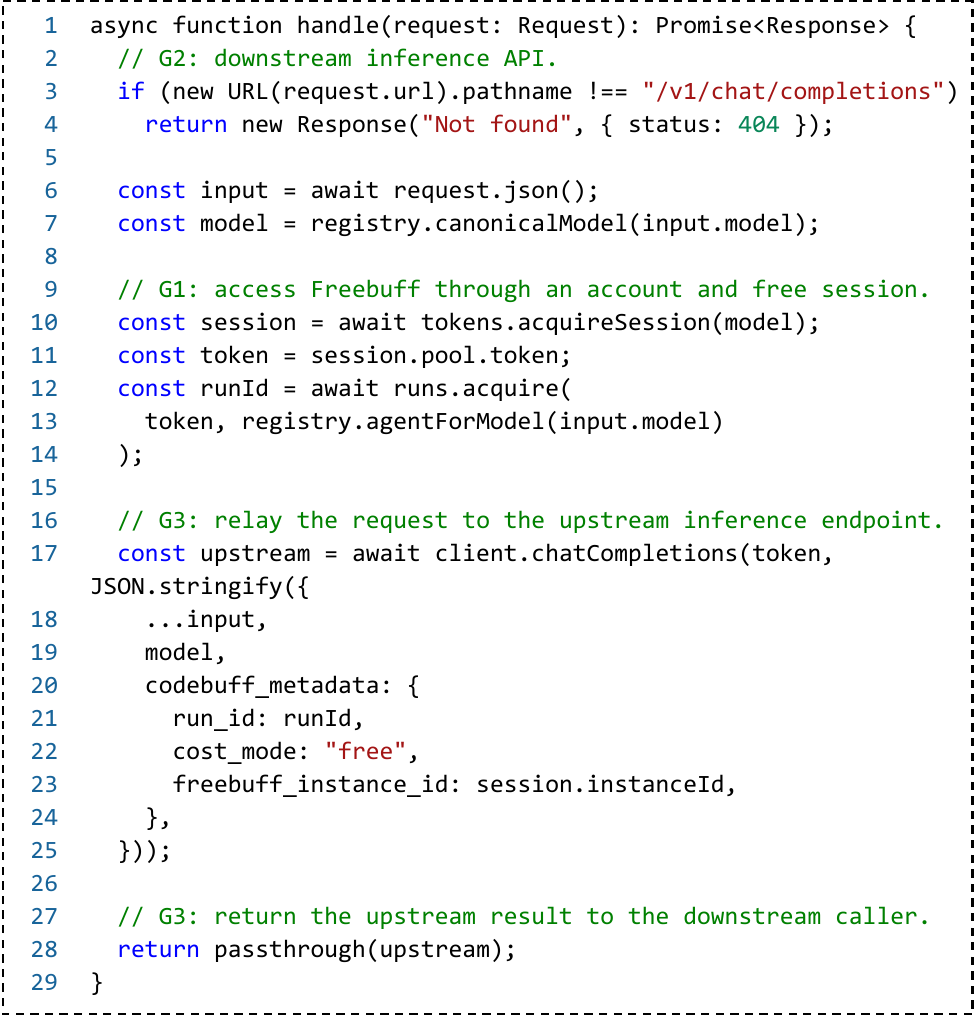}
    \caption{Simplified implementation of a \serv{} project using Freebuff, illustrating G1--G3.}
    \label{fig:example}
\end{figure}

\subsection{Preliminary Study}
\label{subsec:preliminary-study}
We analyze the 76 seed projects to derive features for subsequent project discovery. We first illustrate G1--G3 with a \serv{} project using Freebuff and then describe the feature extraction method applied across the seed set.

\subsubsection{Illustrative example}
\label{subsubsec:seed-example}
\autoref{fig:example} illustrates a qualifying inference path in \href{{https://github.com/Quorinex/Freebuff2API}}{\texttt{Quorinex/Freebuff2API}}. Freebuff offers a free plan that provides inference access through its web application and dedicated clients~\cite{freebuff-plans}.
The project handles requests to \texttt{/v1/chat/completions} (line~3) and reads the input and model selection (lines~6--7), providing the downstream interface required by G2. For a model available through free access, it prepares an upstream session and agent run using an account token (lines~10--14, G1). The call to \texttt{client.chatCompletions} forwards the request to the upstream inference endpoint at \texttt{www.codebuff.com} (line~17). The request includes the run and session identifiers and sets \texttt{cost\_mode} to \texttt{free} (lines~21--23). Finally, \texttt{passthrough} returns the corresponding upstream response to the caller (line~28), completing the inference path required by G3. 

\subsubsection{Feature extraction}
\label{subsubsec:seed-features}
We analyze the 76 seed projects to derive features for locating \serv{} implementations in other repositories and assessing G1--G3. \autoref{tab:seed-characteristics} summarizes the representative features of the seed projects. 

\begin{table}[htbp]
\centering
\small
\caption{Seed characteristics and representative search and analysis features.}
\label{tab:seed-characteristics}
\setlength{\tabcolsep}{4pt}
\resizebox{1\linewidth}{!}{
\begin{tabular}{lrl}
\hline
\textbf{Attribute} & \textbf{\# Proj.} & \textbf{Features} \\
\hline
\rowcolor{blue!6}
\multicolumn{3}{l}{\textbf{Upstream access channels}} \\
Web only & 51 & Domains, endpoints \\
Dedicated client only & 25 & Client identifiers, protocol fields \\
\hline
\rowcolor{blue!6}
\multicolumn{3}{l}{\textbf{Downstream protocols}} \\
OpenAI-compatible & 71 & \texttt{/v1/chat/completions} \\
Anthropic-compatible & 34 & \texttt{/v1/messages} \\
Gemini-compatible & 17 & \texttt{generateContent} \\
Custom & 7 & Custom routes \\
\hline
\rowcolor{blue!6}
\multicolumn{3}{l}{\textbf{Upstream invocation methods}} \\
Direct product requests & 57 & HTTP/WebSocket calls \\
Browser-mediated access & 15 & Browser requests, DOM operations \\
Dedicated-client bridge & 4 & Process invocations \\
\hline
\end{tabular}
}
\end{table}

\noindent\textbf{Upstream services and access channels.}
We first identified the product through which each seed project obtained inference. For network requests, we traced how the adapter constructed the destination URL and extracted the host and product-specific path. For example, although \texttt{Gemini-Web2API} constructs part of its request URL at runtime, the fixed host and path in its URL helper identify Gemini Web as the upstream product. For paths that invoked a dedicated client, we used the client identifier and protocol fields to identify the product. We then consulted the provider's documentation to verify whether the selected model was available free of charge through that channel under the account or session configuration used on the path. Across the seed set, we identified 20 distinct upstream services, whose distribution across projects is shown in \autoref{tab:seed-upstreams}.

\begin{table}[t]
\centering
\caption{Upstream LLM services in the seed set. Services with multiple access channels are grouped under CLI when available.}
\label{tab:seed-upstreams}
\setlength{\tabcolsep}{4pt}
\renewcommand{\arraystretch}{1.1}

\resizebox{\linewidth}{!}{
\begin{tabular}{@{}llr@{\hspace{12pt}}lr@{}}
\toprule
\textbf{Channel}
& \textbf{Upstream service} & \textbf{\# Proj.}
& \textbf{Upstream service} & \textbf{\# Proj.} \\
\midrule
\multirow{7}{*}{Web}
 & Gemini Web       & 15 & DeepSeek Web   & 13 \\
 & Qwen Web         &  6 & Doubao Web     &  4 \\
 & ChatGPT Web      &  3 & Claude Web     &  2 \\
 & Duck.ai          &  2 & Perplexity Web &  2 \\
 & Zhipu Qingyan    &  2 & cto.new        &  1 \\
 & Google AI Studio &  1 & Kimi Web       &  1 \\
 & Metaso AI Search &  1 &                &    \\
\midrule
\multirow{2}{*}{CLI}
 & Google Antigravity & 9 & Kiro          &  5 \\
 & Freebuff           & 5 & Grok Build    &  2 \\
\midrule
\multirow{2}{*}{Desktop}
 & Cline           & 2 & WorkBuddy       &  2 \\
 & iFlytek AStudio & 1 &                 &    \\
\midrule
\multicolumn{4}{@{}l}{\textbf{Total projects (deduplicated)}}
 & \textbf{76} \\
\bottomrule
\end{tabular}
}
\end{table}

\noindent\textbf{Downstream interfaces.}
To assess G2, we followed server-side route registrations and decorators to their handlers, combining route prefixes with local paths where necessary. We identified the protocol from the route and the request and response fields handled there. In \autoref{fig:example}, for instance, line~3 checks the route exposed to downstream callers, while line~17 issues an outbound inference request. Among the 76 seed projects, 71 exposed an OpenAI-compatible interface, 34 an Anthropic-compatible interface, and 17 a Gemini-compatible interface (\autoref{tab:seed-characteristics}). Seven projects exposed custom interfaces, six of which also supported at least one of the three common protocols. We therefore used handlers for these three protocol families as starting points for tracing inference requests in candidate repositories.

\noindent\textbf{Inference paths.}
To derive features for assessing G3, we examined how seed projects connect downstream requests to upstream inference and return the corresponding results. We identified three types of upstream operations: direct HTTP or WebSocket calls, browser-mediated requests or DOM interactions, and invocations of dedicated clients. When a handler supported multiple adapters, we examined the branch selected by the model name or configuration. These operations, together with downstream response construction and streaming code, provide reference points for reconstructing inference paths during subsequent project discovery.
In \autoref{fig:example}, the handler reads the request body into \texttt{input} at line~6 and includes its fields in the upstream call at lines~17--18. The result is assigned to \texttt{upstream} and returned through \texttt{passthrough} at line~28. This connects the caller's input and the corresponding inference output under one configuration. For each qualifying seed path, we recorded the configuration, downstream handler, upstream operation, and supporting source locations.

\subsection{Project Discovery}
\label{subsec:project-discovery}
\noindent\textbf{Data collection.}
We searched GitHub repositories using 24 queries related to free LLM access, web-to-API conversion, and inference proxies (e.g., \texttt{web to llm api} and \texttt{free llm proxy}). Across all queries, we collected 5,963 search results corresponding to 5,181 distinct repositories after deduplication by repository URL. We inspected the contents of these repositories and excluded 149 that provided only curated resource lists (e.g., awesome lists and project collections) or tutorials and contained no implementation of an inference API. The remaining 5,032 repositories formed the candidate set for subsequent source-code analysis.

\textit{(1) Identifying downstream inference APIs (G2).}
We screened the 5,032 candidate repositories for downstream inference APIs compatible with OpenAI, Anthropic, or Gemini, the three most common protocol families in our seed set (\autoref{tab:seed-characteristics}). Using the route patterns in the table, we located server-side route definitions with Tree-sitter and inspected their handlers. We retained a repository if at least one handler accepted inference requests from downstream clients and returned or streamed responses in the corresponding protocol format. This screening yielded 1,216 repositories.

\textit{(2) Tracing end-to-end inference paths (G3).}
For the 1,216 repositories retained under G2, we used Codex with GPT-6 Luna to reconstruct inference paths from source code. Starting from the downstream handlers identified with Tree-sitter, we asked Codex to determine where each handler sent the caller's input for inference and how the resulting output returned to that caller. When a handler supported multiple upstream adapters, Codex followed the branch selected by the model name or configuration. A path satisfied G3 if the caller's input reached an upstream inference operation and the output of that same operation was returned or streamed to the caller under the same configuration. This step retained 767 repositories and excluded 449. 

\textit{(3) Verifying free upstream inference access (G1).}
For each inference path identified in Step~(2), we determined which upstream product and access channel the adapter used by inspecting the endpoint it contacted or the client it invoked. We then consulted the provider's pricing pages and product documentation to verify whether the selected model was available through that web application or dedicated client under a free tier, trial, or promotional grant applicable to the account or session used by the adapter. We retained a repository if at least one of its traced paths met these conditions. Of the 767 repositories examined, 510 contained at least one path satisfying G1--G3.

\textit{(4) Validation and merging with the seed set.}
We manually reviewed all 510 retained repositories and a random sample of 100 excluded repositories. For previously included repositories, we examined the source code supporting the downstream API and its connection to upstream inference, together with the documentation establishing free access. For sampled exclusions, we inspected the implementation for qualifying paths that the screening process may have missed. Manual review confirmed that 507 of the 510 provisionally included repositories satisfied G1--G3, yielding an inclusion precision of 99.4\%. Among the 100 sampled excluded repositories, 2 satisfied the criteria and should have been included, corresponding to a false omission rate of 2\% within the reviewed sample. We corrected the identified misclassifications and merged them with the 76 manually annotated seed projects. After deduplication, the final dataset contained 513 \serv{} projects.

\noindent\textbf{Dataset composition.}
For each retained \serv{} project, we recorded the repository revision, downstream route and handler, adapter and configuration, upstream service and invocation method, and supporting source locations and free-access documentation. Branches with different adapters or request and response implementations were recorded separately. This yielded 513 \serv{} projects and 1,588 inference paths to 64 upstream services. Grouping paths within each project by downstream protocol family, upstream service, and invocation method produced 682 combinations. The individual path records retain endpoint and response-mode differences and support the subsequent analysis.

\noindent\textbf{Limitations.}
Our collection may omit relevant \serv{} projects because discovery depends on public GitHub availability, search terminology, and the three downstream API families considered in this study. Source-code screening and LLM-assisted analysis may also overlook qualifying implementations. We sought to broaden coverage through seed searches across multiple platforms, 24 GitHub queries, and analysis covering direct requests, browser-mediated access, and dedicated-client bridges. To assess screening reliability, we manually reviewed all included repositories and a sample of 100 excluded repositories. These checks support the accuracy of inclusion decisions and help identify omissions among the collected candidates. The final dataset spans 64 upstream services and all three invocation methods, providing a basis for examining LSF mechanisms across diverse services and implementations.
\section{Understanding LLM Service Freeloading}
\label{sec:understanding}

In this section, we characterize the ecosystem of \serv{} projects, identify the providers supplying their inference resources, and analyze the conditions governing the use and redistribution of these resources.

\subsection{Scope and Magnitude}
\label{subsec:scope}

\begin{table*}[t]
\centering
\caption{The ten most-starred repositories in our dataset, with GitHub metrics collected on November 16, 2026.}
\label{tab:top-lsf-projects}
\resizebox{1\linewidth}{!}{
\begin{tabular}{rllrrll}
\hline
\textbf{Repository} & \textbf{Project type} & \textbf{Created}
& \textbf{Stars} & \textbf{Forks} & \textbf{Downstream API}
& \textbf{Upstream service} \\
\hline
\href{https://github.com/Ebony-Vinyl/dsh-our-free-model}
     {\texttt{Ebony-Vinyl/\allowbreak dsh-our-free-model}}
& Client plugin & 2026-09 & 75,841 & 2,127
& OpenAI, Anthropic, Gemini & OpenCode \\

\href{https://github.com/lbjlaq/Antigravity-Manager}
     {\texttt{lbjlaq/\allowbreak Antigravity-Manager}}
& Account manager & 2025-11 & 33,129 & 3,545
& OpenAI, Anthropic, Gemini & Google Antigravity \\

\href{https://github.com/decolua/9router}
     {\texttt{decolua/\allowbreak 9router}}
& Multi-service gateway & 2026-01 & 32,484 & 6,363
& OpenAI, Anthropic, Gemini & DeepSeek Web \\

\href{https://github.com/basketikun/chatgpt2api}
     {\texttt{basketikun/\allowbreak chatgpt2api}}
& API proxy & 2026-04 & 6,905 & 1,746
& OpenAI, Anthropic & ChatGPT Web \\

\href{https://github.com/su-kaka/agy2api}
     {\texttt{su-kaka/\allowbreak agy2api}}
& API proxy & 2025-08 & 5,360 & 1,380
& OpenAI, Anthropic, Gemini & Google Antigravity \\

\href{https://github.com/badrisnarayanan/antigravity-claude-proxy}
     {\texttt{badr***an/\allowbreak antigravity-claude-proxy}}
& API proxy & 2025-12 & 4,062 & 578
& Anthropic & Google Antigravity \\

\href{https://github.com/Sophomoresty/gemini-web2api}
     {\texttt{Sophomoresty/\allowbreak gemini-web2api}}
& API proxy & 2026-05 & 3,598 & 787
& OpenAI, Gemini & Gemini Web \\

\href{https://github.com/aurora-develop/aurora}
     {\texttt{aurora-develop/\allowbreak aurora}}
& API proxy & 2024-04 & 2,595 & 710
& OpenAI & ChatGPT Web \\

\href{https://github.com/jwadow/kiro-gateway}
     {\texttt{jwadow/\allowbreak kiro-gateway}}
& API gateway & 2025-12 & 2,357 & 562
& OpenAI, Anthropic & Kiro \\

\href{https://github.com/hj01857655/kiro-account-manager}
     {\texttt{hj01857655/\allowbreak kiro-account-manager}}
& Account manager & 2025-12 & 2,087 & 361
& OpenAI, Anthropic & Kiro \\
\hline
\end{tabular}}

\end{table*}

\begin{figure}[t]
    \centering
    \includegraphics[width=0.95\linewidth]{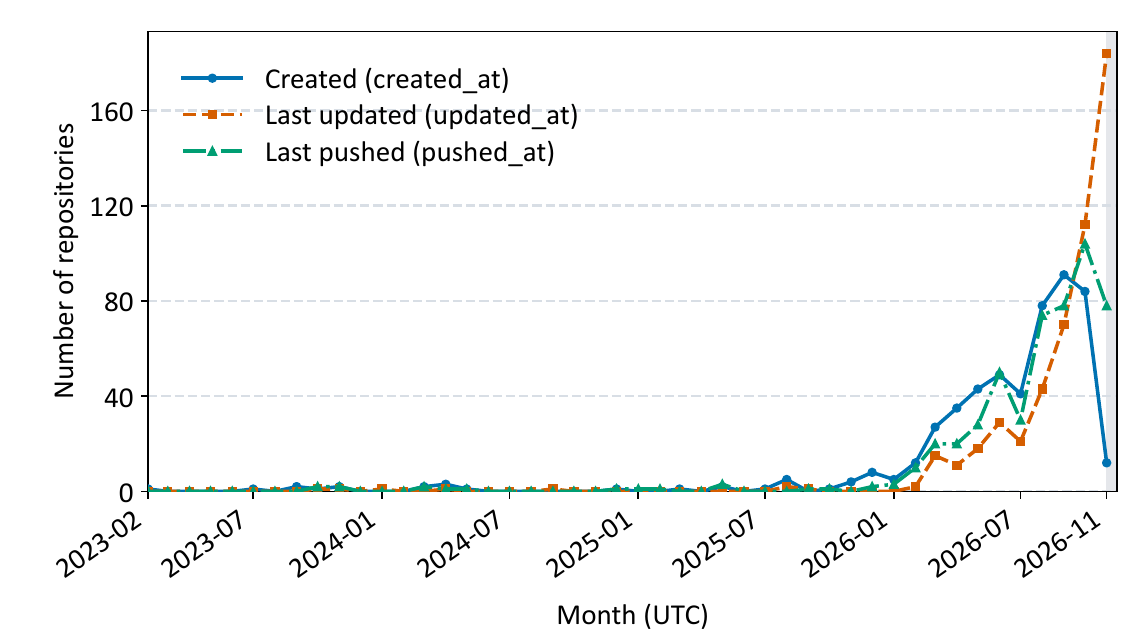}
    \caption{Monthly distribution of creation, last-update, and last-push dates for \serv{} repositories.}
    \label{fig:repository_timestamps}
\end{figure}

\noindent\textbf{Repository growth.}
Our dataset is dominated by recently created repositories: 477 of the 513 projects (93.0\%) were created in 2026. As shown in \autoref{fig:repository_timestamps}, monthly creation nearly doubled from 41 in July to 78 in August, reached 91 in September, and remained high at 84 in October. These three months accounted for 253 repositories (49.3\% of the dataset), indicating sustained expansion. November was still in progress at the time of submission, so its repository count reflects only a partial month and is excluded from comparisons with complete calendar months.
The last-update and last-push series record each repository's latest timestamp, so their concentration near the snapshot does not establish an increase in development frequency. We examined recent pushes separately for new and established repositories. Of the 286 repositories with a push during the preceding 90 days, 211 (73.8\%) were created within that window. Among the 302 repositories that predated the window, only 75 (24.8\%) received a push. Thus, much of the recent push activity accompanied repository creation. Commit histories provide a further distinction between importing existing code and developing it after creation.

\begin{figure}[t]
    \centering
    \includegraphics[width=0.95\linewidth]{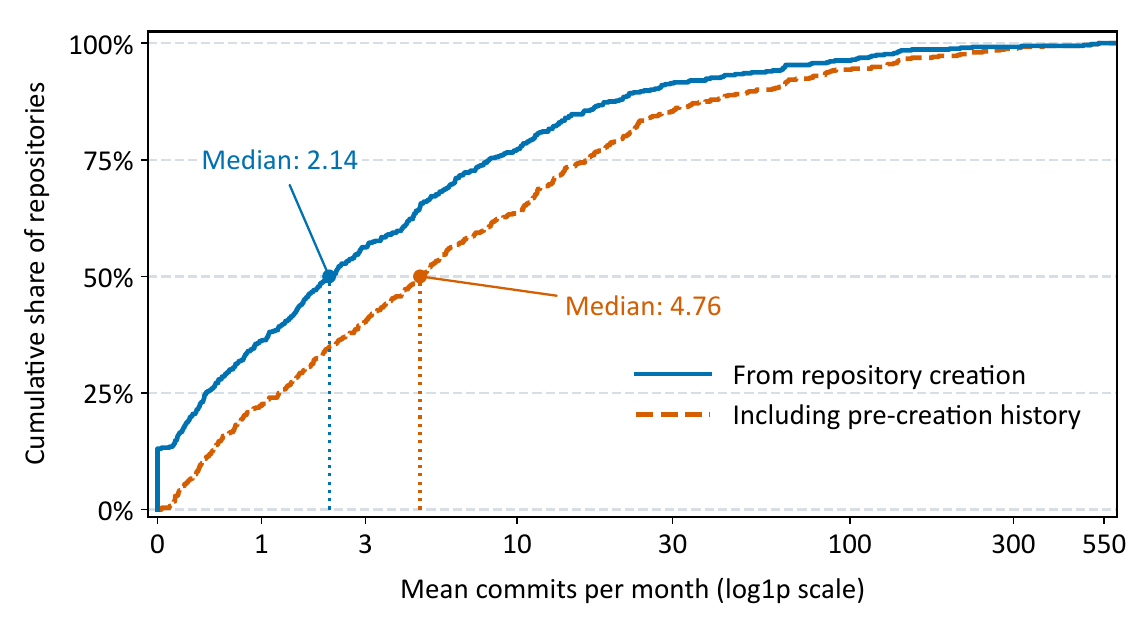}
    \caption{Monthly commit frequency of \serv{} repositories with and without pre-creation history.}
    \label{fig:lsf_commit_frequency}
\end{figure}

\noindent\textbf{Code reuse and development activity.} We further compared commit dates with repository creation dates on the default branch. We found that 283 contained earlier commits. Counting them more than doubled the median monthly commit rate, from 2.14 to 4.76, as shown in \autoref{fig:lsf_commit_frequency}. Across 90 repositories, we identified 1,828 distinct commits shared by two or more repositories. For example, \href{https://github.com/wsl-fd/deepseek-free-api1}{\texttt{wsl-fd/deepseek-free-api1}} contained all 48 commits found in \href{https://github.com/jonnyquan/deepseek-free-api}{\texttt{jonnyquan/deepseek-free-api}}. Although the former had a longer history (66 commits), only three commits followed its creation, compared with 45 in the latter. To assess whether development continued, we followed 199 repositories with a commit in their creation month and three full months of subsequent observation. Of these, 110 (55.3\%) had no commits in those three months. Across the full sample of 513 repositories, ten accounted for 41.5\% of post-creation commits. Two of them, \href{https://github.com/decolua/9router}{\texttt{decolua/9router}} and \href{https://github.com/su-kaka/agy2api}{\texttt{su-kaka/agy2api}}, recorded commits in every observed calendar month over 11 and 16 months, respectively.

\noindent\textbf{Project popularity.}
To assess community interest in \serv{} projects, we analyzed the distribution of GitHub stars across our dataset. We found that attention was concentrated in a small subset of projects: the median repository had one star, and 383 of the 513 repositories (74.7\%) had fewer than ten, while the ten projects in \autoref{tab:top-lsf-projects} accounted for 89.0\% of all stars. These include \texttt{decolua/9router} and \texttt{su-kaka/agy2api}, which also maintained the continuous development activity described above.
We also found that the most popular projects incorporate free inference access into broader integration software. The three leading repositories provide a client plugin, an account manager, and a multi-service gateway, respectively, and together accounted for 74.8\% of all stars. Eight of these projects support multiple downstream protocols. These projects make free inference available through interfaces already supported by existing applications, while handling account access and service-specific interactions internally. 

\subsection{Affected LLM Services}

\noindent\textbf{Upstream services and free access.}
Our review of 513 \serv{} projects identified 64 upstream LLM services with documented free access. As shown in \autoref{tab:free-access-types}, we identified three forms of free access. 51 services offered a free tier, 24 offered a trial, and 18 granted promotional credits. These categories overlap. A \emph{free tier} provides ongoing access to inference, often with an allowance that renews periodically. Kiro Free, for example, provides 50 credits per month~\cite{kiro-pricing}. A \emph{free trial} lets users try a service or paid feature for a limited time or until they exhaust a fixed allowance. Notion AI gives users on its \emph{Free} and \emph{Plus} plans a limited number of complimentary responses, after which they must upgrade to continue using the feature~\cite{notion-ai-trial}. \emph{Promotional credits} are awarded for a qualifying action, such as signing up or referring another user. WorkBuddy's international edition illustrates how a service can combine a free tier with a promotion. Its free plan provides 100 credits per month, and new users receive an additional 250 credits at first sign-in that expire after 14 days~\cite{workbuddy-subscription}.

\begin{table}[t]
\centering
\caption{Project coverage of upstream LLM services by form of free access.}
\label{tab:free-access-types}
\resizebox{1\linewidth}{!}{
\begin{tabular}{crrl}
\hline
\textbf{Access type} & \textbf{\# Services} & \textbf{\# Projects}
& \textbf{Projects / service}\textsuperscript{a}  \\
\hline
Free tier              & 51 & 404 & 2 (1--105) \\
Free trial             & 24 & 209 & 3 (1--86)  \\
Promotional credits    & 18 &  49 & 2 (1--13)  \\
\hline
\textbf{Total (deduplicated) }         & 64 & 513 & 2 (1--105) \\
\hline
\end{tabular}}

\smallskip
\begin{minipage}{\linewidth}
\footnotesize
\textit{Note.} \textsuperscript{a}\,Median number of projects per service (minimum--maximum). 
\end{minipage}
\end{table}

\noindent\textbf{Concentration across services.}
We further examined how the 513 \serv{} projects used the 64 upstream services. As shown in \autoref{tab:free-access-types}, 404 projects supported a service offering a free tier, 209 supported one offering a free trial, and 49 supported one offering promotional credits. The median number of projects per service ranged from two to three across these categories. We found that 264 of the 513 projects (51.5\%) supported at least one of four web products: DeepSeek Web (105), Gemini Web (86), Qwen Web (57), and ChatGPT Web (37). DeepSeek Web was the most widely supported of the four. In April 2026, the maintainer of \texttt{ds2api} raised concerns about operators registering DeepSeek accounts in bulk and using the project to relay requests for others~\cite{ds2api-issue344}.

\begin{figure}[t]
    \centering
    \includegraphics[width=1\linewidth]{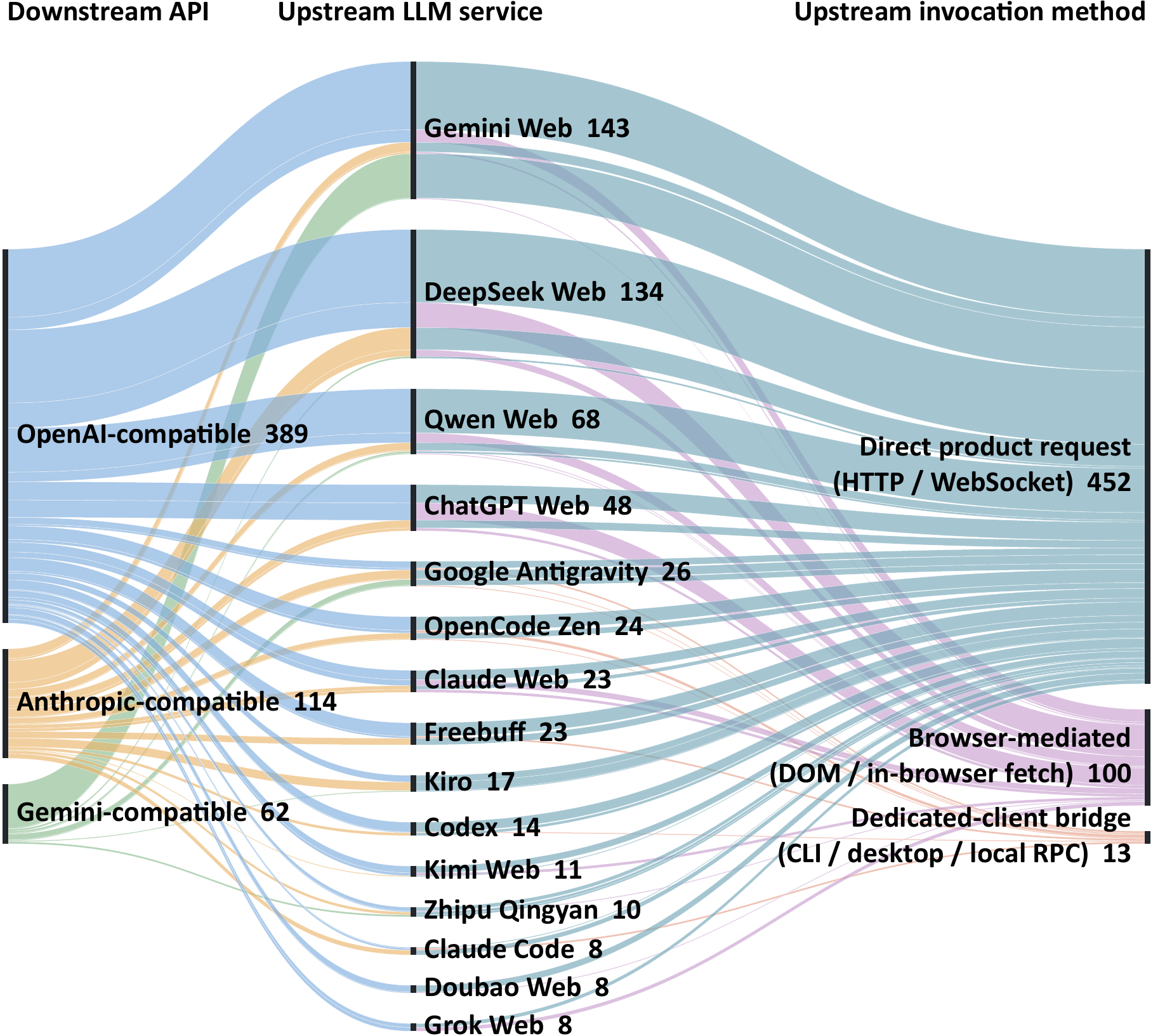}
    \caption{Paths from downstream APIs to the top 15 upstream LLM services, grouped by invocation method.}
    \label{fig:lsf_sankey}
\end{figure}

\noindent\textbf{Access paths to upstream services.}
To characterize how downstream API requests reach these services, we traced 1,588 inference paths across 513 \serv{} projects.  The paths connected OpenAI-, Anthropic-, or Gemini-compatible APIs to 64 upstream LLM services with documented free access. Within each project, we counted paths once if they shared the same downstream protocol family, upstream service, and invocation method, even when they used different endpoints (e.g., \texttt{/v1/chat/completions} vs.\ \texttt{/v1/responses}) or response formats (e.g., non-streaming JSON vs.\ streaming SSE). This yielded 682 distinct combinations. \autoref{fig:lsf_sankey} shows the 15 upstream services that accounted for the most combinations (565 in total). 
We distinguished three invocation methods according to which component submits the inference request. 452 combinations used a \emph{direct product request}: its own code reproduced the HTTP or WebSocket request normally sent by the product client. These projects could therefore serve downstream requests without running a browser or invoking a dedicated client for each inference call. In 100 combinations, the project used \emph{browser-mediated access}, either interacting with the product's web page or issuing a request from within the browser session. The remaining 13 used a \emph{dedicated-client bridge}, passing the prompt to a command-line or desktop client, sometimes through a local RPC interface, and relaying the client's response downstream.

\begin{table*}[t]
\centering
\small
\setlength{\tabcolsep}{4pt}
\caption{Provider and service restrictions relevant to \serv{}. AA: automated access; AS: account sharing; SR: service redistribution; MA: multiple accounts and free allowances; LC: limit circumvention.}
\label{tab:provider-policy-restrictions}
\begin{minipage}[t]{.494\linewidth}
\vspace{0pt}
\centering
\resizebox{\linewidth}{!}{
\begin{tabular}{llccccc}
\hline
\textbf{Provider} & \textbf{LLM Service} & \textbf{AA} & \textbf{AS} & \textbf{SR} & \textbf{MA} & \textbf{LC} \\
\hline
Aelius & \href{https://aeliusai.com/usage}{Aelius Web Chat}
& \yes & \yes & \partialcell & \yes & \yes \\
\hline
Aethyro & \href{https://aethyro.com/terms.html}{Aethyro}
& \no & \no & \no & \no & \no \\
\hline
\multirow{2}{*}{Alibaba} & \href{https://qwen.ai/termsservice?language=en-US}{Qwen Code (OAuth)}
& \yes & \yes & \yes & \no & \yes \\
 & \href{https://qwen.ai/termsservice?language=en-US}{Qwen Web (Intl.)}
& \yes & \yes & \yes & \no & \yes \\
\hline
Anomaly & \href{https://opencode.ai/legal/terms-of-service}{OpenCode Zen}
& \yes & \partialcell & \yes & \yes & \yes \\
\hline
\multirow{2}{*}{Anthropic} & \href{https://code.claude.com/docs/en/legal-and-compliance}{Claude Code}
& \yes & \yes & \yes & \partialcell & \yes \\
 & \href{https://www.anthropic.com/legal/consumer-terms}{Claude Web}
& \yes & \yes & \yes & \partialcell & \yes \\
\hline
Anysphere & \href{https://cursor.com/terms-of-service}{Cursor}
& \yes & \no & \yes & \no & \no \\
\hline
Arena Intelligence & \href{https://help.arena.ai/articles/5629909088-terms-of-use}{Arena}
& \yes & \yes & \yes & \yes & \yes \\
\hline
AWS & \href{https://aws.amazon.com/agreement/}{Kiro}
& \partialcell & \partialcell & \partialcell & \partialcell & \yes \\
\hline
Baidu & \href{https://ernie.baidu.com/web/src/pages/webuserprivacy/index.html}{Baidu Chat}
& \yes & \yes & \partialcell & \partialcell & \yes \\
\hline
Bright Zenith & \href{https://qoder.com/product-service}{Qoder}
& \yes & \yes & \no & \partialcell & \yes \\
\hline
\multirow{2}{*}{ByteDance} & \href{https://www.doubao.com/legal/terms}{Doubao Web (China)}
& \partialcell & \yes & \partialcell & \partialcell & \no \\
 & \href{https://www.trae.cn/terms-of-service}{TRAE SOLO (CN)}
& \yes & \yes & \yes & \partialcell & \yes \\
\hline
Chipotle & \href{https://www.chipotle.com/terms-of-use}{Chipotle Pepper}
& \no & \partialcell & \yes & \no & \no \\
\hline
Cline Bot & \href{https://cline.bot/tos}{Cline}
& \partialcell & \yes & \yes & \no & \yes \\
\hline
DappHero & \href{https://freepi.ai/terms}{FreePi}
& \partialcell & \yes & \yes & \partialcell & \yes \\
\hline
DeepSeek & \href{https://cdn.deepseek.com/policies/en-US/deepseek-terms-of-use.html}{DeepSeek Web}
& \yes & \yes & \partialcell & \partialcell & \no \\
\hline
DuckDuckGo & \href{https://duckduckgo.com/duckai/privacy-terms}{Duck.ai}
& \yes & \no & \yes & \no & \yes \\
\hline
Engine Labs & \href{https://www.iubenda.com/terms-and-conditions/55632616}{cto.new}
& \no & \no & \partialcell & \no & \yes \\
\hline
Errante Group & \href{https://typli.ai/legal/terms-and-conditions}{Typli.ai}
& \yes & \yes & \no & \yes & \no \\
\hline
Freebuff & \href{https://freebuff.com/terms-of-service}{Freebuff}
& \yes & \yes & \yes & \yes & \yes \\
\hline
GitHub & \href{https://docs.github.com/en/site-policy/github-terms/github-terms-of-service}{Copilot (individual account)}
& \partialcell & \yes & \partialcell & \yes & \partialcell \\
\hline
\multirow{5}{*}{Google} & \href{https://geminicli.com/docs/resources/tos-privacy/}{Gemini CLI}
& \yes & \no & \partialcell & \no & \yes \\
 & \href{https://policies.google.com/terms}{Gemini Web (US)}
& \partialcell & \no & \yes & \no & \yes \\
 & \href{https://ai.google.dev/gemini-api/terms}{AI Studio (developer)}
& \partialcell & \yes & \yes & \no & \yes \\
 & \href{https://antigravity.google/terms}{Antigravity (US consumer)}
& \yes & \no & \yes & \no & \yes \\
 & \href{https://policies.google.com/terms}{Google Search AI Mode}
& \partialcell & \no & \partialcell & \no & \yes \\
\hline
Huawei & \href{https://www.huaweicloud.com/declaration/tsa-codearts-doer.html}{CodeArts Agent (CN)}
& \yes & \no & \partialcell & \no & \no \\
\hline
Hugging Face & \href{https://huggingface.co/terms-of-service}{HuggingChat}
& \partialcell & \yes & \no & \no & \yes \\
\hline
JetBrains & \href{https://www.jetbrains.com/legal/docs/terms/jetbrains-ai-service/}{JetBrains AI}
& \no & \no & \partialcell & \no & \yes \\
\hline
Langbase & \href{https://commandcode.ai/terms}{CommandCode}
& \yes & \no & \yes & \yes & \yes \\
\hline
\end{tabular}}
\end{minipage}
\hfill
\begin{minipage}[t]{.494\linewidth}
\vspace{0pt}
\centering
\resizebox{\linewidth}{!}{
\begin{tabular}{llccccc}
\hline
\textbf{LLM Provider} & \textbf{LLM Service} & \textbf{AA} & \textbf{AS} & \textbf{SR} & \textbf{MA} & \textbf{LC} \\
\hline
Logicc & \href{https://www.logicc.com/en/agb}{Logicc}
& \no & \yes & \partialcell & \no & \partialcell \\
\hline
MainFunc & \href{https://www.genspark.ai/terms}{Genspark Web}
& \yes & \yes & \no & \no & \no \\
\hline
Meituan & \href{https://www.tabbit.com/terms}{Tabbit (China)}
& \partialcell & \partialcell & \partialcell & \partialcell & \partialcell \\
\hline
\multirow{2}{*}{Meta} & \href{https://www.facebook.com/legal/ai-terms}{Meta AI Web}
& \partialcell & \partialcell & \no & \partialcell & \yes \\
 & \href{https://muse.ai/terms}{Muse}
& \partialcell & \partialcell & \no & \partialcell & \yes \\
\hline
Metaso & \href{https://metaso.cn/meta-user-policy}{Metaso AI Search}
& \yes & \partialcell & \yes & \partialcell & \no \\
\hline
Microsoft & \href{https://www.microsoft.com/en-us/servicesagreement}{Copilot Web (consumer)}
& \yes & \yes & \no & \partialcell & \yes \\
\hline
MiniMax & \href{https://code.minimaxi.com/doc/zh/terms-of-service.html}{MiniMax Agent (China)}
& \partialcell & \yes & \partialcell & \partialcell & \no \\
\hline
Mistral AI & \href{https://legal.mistral.ai/terms/row-consumer-terms/}{Le Chat (non-EEA consumer)}
& \partialcell & \yes & \yes & \yes & \yes \\
\hline
Moonshot AI & \href{https://www.kimi.com/user/agreement/modelUse?version=v2}{Kimi Web (China)}
& \yes & \yes & \partialcell & \partialcell & \no \\
\hline
Notion & \href{https://notion.notion.site/Notion-AI-Supplementary-Terms-fa9034c8b5a04818a6baf3eac2adddbb}{Notion AI (personal)}
& \partialcell & \yes & \no & \no & \partialcell \\
\hline
NVIDIA & \href{https://developer.nvidia.com/legal/terms}{NVIDIA Playground}
& \partialcell & \yes & \partialcell & \no & \yes \\
\hline
OmniRush & \href{https://omnirush.ai/terms}{OmniRush}
& \partialcell & \yes & \partialcell & \yes & \partialcell \\
\hline
\multirow{3}{*}{OpenAI} & \href{https://openai.com/policies/terms-of-use/}{ChatGPT Web}
& \yes & \yes & \yes & \no & \yes \\
 & \href{https://help.openai.com/en/articles/11369540-using-codex-with-your-chatgpt-plan}{Codex (ChatGPT plan)}
& \yes & \yes & \yes & \no & \yes \\
 & \href{https://openai.com/policies/row-terms-of-use/}{Prism (personal)}
& \yes & \yes & \yes & \no & \yes \\
\hline
Perplexity & \href{https://www.perplexity.ai/en-GB/hub/legal/terms-of-service}{Perplexity Web}
& \partialcell & \no & \partialcell & \partialcell & \partialcell \\
\hline
StepFun & \href{https://chat.stepfun.com/legal/terms}{StepChat / StepFun}
& \partialcell & \yes & \partialcell & \partialcell & \no \\
\hline
Superdesign & \href{https://superdesign.dev/}{Superdesign}
& \multicolumn{5}{c}{\textit{Policy unavailable}} \\
\hline
\multirow{5}{*}{Tencent} & \href{https://rule.tencent.com/rule/202607160004}{CodeBuddy (China, personal)}
& \partialcell & \yes & \partialcell & \partialcell & \partialcell \\
 & \href{https://www.codebuddy.ai/document/term}{CodeBuddy (Intl.)}
& \yes & \no & \yes & \no & \yes \\
 & \href{https://rule.tencent.com/rule/202603180001}{WorkBuddy (China, personal)}
& \partialcell & \yes & \partialcell & \partialcell & \partialcell \\
 & \href{https://www.workbuddy.ai/document/term}{WorkBuddy (Intl.)}
& \yes & \no & \yes & \no & \yes \\
\hline
uhelp.ai & \href{https://www.uhelp.ai/}{uhelp.ai}
& \multicolumn{5}{c}{\textit{Policy unavailable}} \\
\hline
\multirow{2}{*}{xAI} & \href{https://x.ai/legal/terms-of-service}{Grok Build (consumer)}
& \partialcell & \yes & \yes & \no & \yes \\
 & \href{https://x.ai/legal/terms-of-service}{Grok Web}
& \yes & \yes & \yes & \partialcell & \yes \\
\hline
Xiaomi & \href{https://aistudio-cdn.xiaomimimo.com/xiaomimimo-static/policy/agreement/zh_CN.html}{MiMo Studio (Chinese terms)}
& \partialcell & \partialcell & \partialcell & \partialcell & \no \\
\hline
YTL AI Labs & \href{https://www.ilmu.ai/legal/terms-of-service/}{ILMUchat}
& \partialcell & \yes & \yes & \partialcell & \yes \\
\hline
\multirow{4}{*}{Zhipu} & \href{https://autoclaw.zhipuai.cn/privacy/md2html/?favicon=autoglm\&md=autoclaw_agreement}{AutoClaw (CN)}
& \yes & \yes & \partialcell & \partialcell & \yes \\
 & \href{https://z-cdn.chatglm.cn/z-ai/static/Z.ai\%E7\%94\%A8\%E6\%88\%B7\%E5\%8D\%8F\%E8\%AE\%AE.html}{Z.ai Chat (Chinese terms)}
& \yes & \partialcell & \no & \partialcell & \yes \\
 & \href{https://zcode.z.ai/en/terms}{ZCode}
& \yes & \yes & \no & \partialcell & \yes \\
 & \href{https://chatglm.cn/legal/appAgreement}{Qingyan (China, app)}
& \yes & \partialcell & \no & \no & \yes \\
\hline
\end{tabular}}
\end{minipage}
\smallskip
\begin{minipage}{\linewidth}
\footnotesize
\textit{Note.} 
\yes\ prohibited; \partialcell\ conditional; \no\ not stated, not permission. 
\textit{Policy unavailable}: we could not retrieve the agreement text from Superdesign's terms page; uhelp.ai displayed a Terms label in its footer, but we could not obtain the corresponding agreement text. 
\end{minipage}
\end{table*}

\subsection{Provider Restrictions on \serv{}}
To determine which provider restrictions apply to \serv{}, we reviewed the terms governing each upstream product. We started with the terms linked from the product's official site or sign-up page and followed references to product-specific agreements and acceptable use policies. For each service, we checked which edition, region, and account type the documents covered. We included provider-wide terms when their stated scope covered the product under study. For a free trial or credit grant, we also consulted the offer terms for restrictions on eligibility or repeated claims.
We selected provisions according to the conduct they governed. An \serv{} operator accesses an upstream product and makes its inference capability available to downstream users. This raises separate questions about the means of access (AA), who may use the account or credentials (AS), and whether the operator may provide the service to others (SR). We also examined restrictions on obtaining additional free allowances (MA) and evading limits on their use (LC). For each relevant provision, we recorded the source and section and read it together with the definitions, conditions, and exceptions that determined its scope.

\begin{itemize}[leftmargin=15pt]
    \item \textbf{Automated access (AA).} Use of scripts, browser automation, or unofficial clients to submit prompts to the product or extract its responses.
    \item \textbf{Account sharing (AS).} Allowing another party to use an upstream account, including by lending the account or sharing a session cookie or access token.
    \item \textbf{Service redistribution (SR).} Providing upstream inference to other users or applications through a relay, either free of charge or for payment. The relay operator may retain sole control of the upstream account.
    \item \textbf{Multiple accounts and free allowances (MA).} Creating or using additional accounts, or repeatedly claiming benefits such as free trials and sign-up credits.
    \item \textbf{Limit circumvention (LC).} Evading a provider's usage limits or access controls, for example by rotating accounts to bypass a per-person quota or misrepresenting location to access a region-restricted model.
\end{itemize}

We coded each service and practice separately. \emph{Prohibited} denotes a clause that directly bars the practice for users without special authorization. A written-permission exception is consistent with this code. \emph{Conditional} denotes a prohibition limited to a particular purpose or circumstance. For example, a ban on providing the service to third parties covers both free and paid relays, whereas a ban limited to commercial resale covers only the latter. We used \emph{not stated} when we reviewed the applicable terms and found no provision covering the practice.  We recorded \emph{policy unavailable} separately when we could not obtain the applicable text.

\noindent\textbf{Restrictions analysis.}
As shown in \autoref{tab:provider-policy-restrictions}, 58 of the 62 upstream services with available terms explicitly prohibited at least one practice associated with \serv{}. The terms for 50 services restricted redistribution, either outright or under specified conditions. Of the 12 services whose terms did not address redistribution, 11 explicitly prohibited at least one other practice we examined. For example, the terms reviewed for Qoder and Z.ai Chat did not address redistribution, but both prohibited automated access and limit circumvention. For a relay built on either product, how it submits requests and handles an exhausted quota therefore matters even when the terms say little about the downstream API itself. The scope of redistribution restrictions also varied across editions of the same product. For both CodeBuddy and WorkBuddy, the international terms expressly prohibited redistribution, while the Chinese personal editions restricted it under specified conditions. 
Overall, repackaging free inference as a downstream API commonly conflicts with provider terms, either because the terms restrict redistribution itself or because they prohibit the means used to obtain upstream access. Yet providers vary in how directly they address \serv{}, and some leave the status of third-party relays unspecified. This gap motivates our analysis in \autoref{sec:mechanism} of how \serv{} projects repurpose consumer access and extend free allowances in practice, so providers can better identify and address the abuse their services face.

\section{\serv{} Mechanism Analysis}
\label{sec:mechanism}

This section describes the mechanisms that connect downstream requests to upstream services and analyzes their prevalence across the collected paths.

\begin{figure*}[t]
    \centering
    \includegraphics[width=0.95\linewidth]{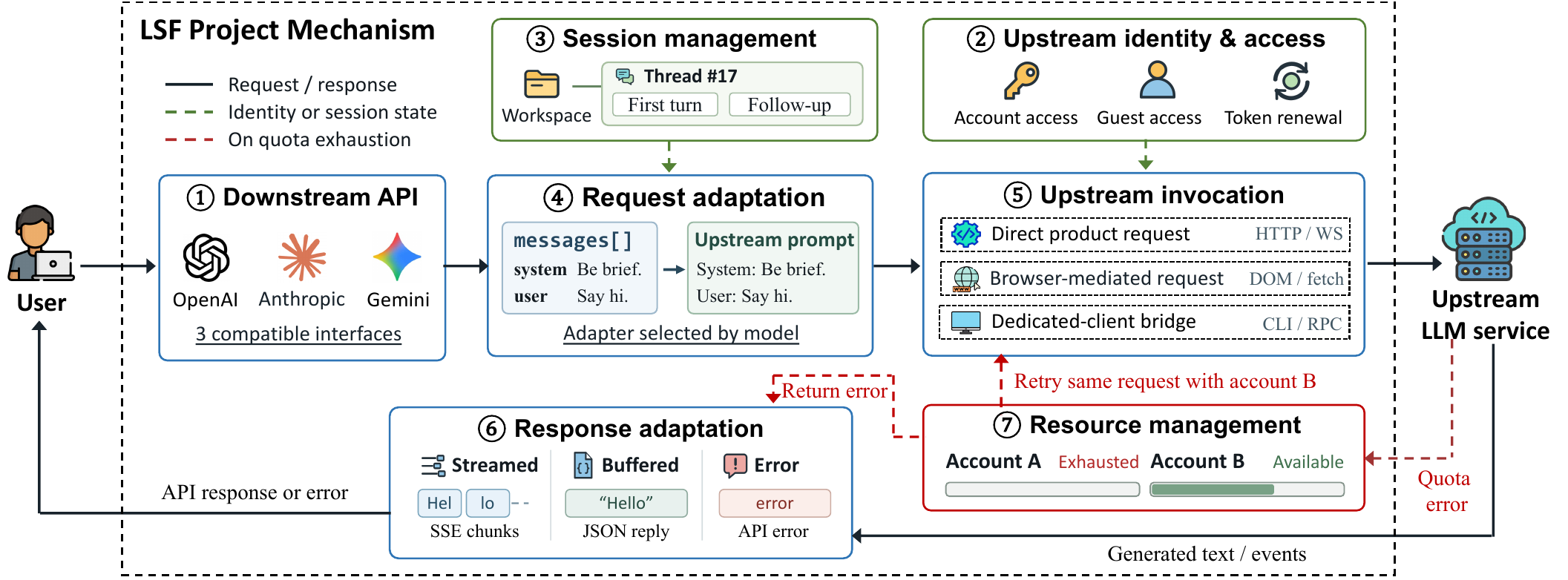}
    \caption{Mechanisms used by \serv{} projects to serve downstream API requests through free access to upstream LLM services. The circled numbers identify functional groups. The arrows show the main request and response path, from the downstream API~\ding{172} through request adaptation~\ding{175} and upstream invocation~\ding{176} to response adaptation~\ding{177}. Upstream identity and access~\ding{173} and session management~\ding{174} supply credentials and conversation state as needed. On quota exhaustion, resource management~\ding{178} may return an error or retry the request under another account.}
    \label{fig:lsf_mechanism}
\end{figure*}

\subsection{Analysis Methodology}

We analyze the 1,588 inference paths identified in \autoref{subsec:project-discovery} to characterize how \serv{} projects obtain and deliver upstream inference. The analysis uses the repository revision and configuration recorded for each path and includes supporting code for credential handling, session management, account scheduling, and error recovery.

\noindent\textbf{Coding scheme.}
We randomly sample 50 of the 513 projects without replacement and include all 182 inference paths recorded for them. The sample contains 156 direct, 20 browser-mediated, and 6 client-mediated paths across 18 upstream services. Two authors independently inspect each path at its recorded revision and configuration, following the downstream input through upstream submission and response delivery. We also inspect supporting functions that provide credentials, prepare product sessions, select accounts, or handle failures. Each researcher describes the implemented behaviors and records their source locations, execution conditions, and connection to the inference path.
The initial inspection distinguishes behaviors within the same function. For example, Antigravity-Tools-LS maps the requested model, checks its recorded allowance, and selects account credentials within one routing function. These behaviors concern request adaptation, quota checking, and account selection, respectively. On the return path, we distinguish extracting generated text, converting response fields, and accumulating streamed output.
We compare our descriptions, inspect omitted or disputed behaviors, and consolidate equivalent implementations into 16 initial mechanism types. Each type receives a definition, assignment criteria, and source examples. These types are provisionally organized into seven groups according to their roles: \textit{downstream API handling}, \textit{upstream identity and access}, \textit{session and workflow management}, \textit{request adaptation}, \textit{upstream invocation}, \textit{response adaptation}, and \textit{resource management}. Four groups describe the main request and response flow; the other three supply product identity, maintain product state, and coordinate account availability and execution. 

\noindent\textbf{Path annotation and agreement.}
Using the initial guide, both authors independently annotate every sampled path without consulting each other's assignments. Each assignment links a mechanism to source locations and the conditions under which it participates in that path. Shared helpers are assigned only to paths that invoke them under the recorded configuration. The guide also distinguishes related behaviors: JSON deserialization alone does not establish request validation; refreshing a credential does not establish account rotation; and forwarding native tool-call fields does not establish tool-call emulation. Account failover requires a connection between the triggering failure, selection of another account, and resubmission of the pending request.
A path may receive multiple mechanism labels, with each mechanism counted once per path. Before resolving disagreements, the two authors agreed on 92\% of path annotations across mechanisms. The authors examine disagreements against the source code, distinguishing missed implementations from differing interpretations of the assignment criteria. Revised definitions are independently reapplied to affected paths before the assignments are reconciled. 

\noindent\textbf{Dataset annotation.}
The resulting guide is applied to the remaining 463 projects using Codex with GPT-6 Luna. Given a path's revision, configuration, and request and response trace, Codex proposes mechanism labels with source references and explanations. Two authors independently review these proposals, inspect the relevant code, and check for omitted mechanisms. Of the 1,398 proposed path--mechanism assignments, manual review removes 23 incorrect assignments and adds 8 omitted assignments. These counts are recorded after the authors reconcile their reviews. Implementations outside the existing definitions are examined before introducing or revising a mechanism type, and each change is applied to previously reviewed paths. Changes resulting from revisions to the coding guide are tracked separately from annotation errors. When unavailable dependencies or deployment settings prevent a decision, the missing information is recorded and the affected label is withheld; these cases are reported separately from agreement calculations. The final scheme contains the 30 mechanisms in \autoref{tab:lsf-mechanism-counts}, with all 1,588 paths manually reviewed.

\subsection{\serv{} Mechanism}
\subsubsection{Operational mechanisms}

We organize the mechanisms implemented by \serv{} projects into seven groups according to their role in serving an inference request, as shown in \autoref{fig:lsf_mechanism}. The first six connect downstream request handling to upstream inference and response delivery; resource management coordinates accounts and execution across this path. In the configurations examined, these mechanisms allow an independent API to serve inference obtained through documented free product access. Credentials, sessions, and account state may persist across requests.

\noindent\textbf{Access control.}
\serv{} projects manage access at two interfaces. Downstream API handling (\ding{172}) accepts and validates requests and may authenticate users with a locally managed API key. Upstream identity and access (\ding{173}) supplies the identity required by the product, either an anonymous session or a registered account. Account credentials may be obtained through login, imported as cookies or tokens, and renewed when they expire. Separating these interfaces lets the operator determine who can use the downstream API while retaining control of the upstream credentials. Several downstream users can consequently consume inference under the same product identity without receiving its credentials, while upstream authorization and allowance accounting remain associated with that identity.

\noindent\textbf{Upstream request construction.}
LSF inference may depend on a conversation thread, workspace, or agent run. Session and workflow management (\ding{174}) establishes this context and determines whether successive requests create new state or reuse an existing session. Request adaptation (\ding{175}) then translates each downstream request into an operation the product accepts. This includes mapping model names to upstream models or agents, converting messages and generation parameters, and attaching the relevant session or task identifiers. Some implementations preserve structured messages; others reconstruct a prompt from conversation history or upload attachments separately and reference them in the inference request. These mechanisms reproduce the preparation normally performed by the product interface, allowing downstream applications to initiate inference without managing the product's workflow themselves.

\noindent\textbf{Upstream invocation and response delivery.}
Upstream invocation (\ding{176}) takes three forms, distinguished by the component that interacts with the product. With \emph{direct product requests}, the project submits inference through its own HTTP or WebSocket client. \emph{Browser-mediated} implementations operate the product's web interface or issue requests within a browser session. A \emph{dedicated-client bridge} delegates execution to a product component, such as a CLI, desktop client, or Language Server, through its local interfaces. These methods provide different ways to make product inference available behind a downstream API.
Response adaptation (\ding{177}) completes this connection. Compatible responses may pass through directly, while other implementations translate message structures, reconstruct streaming events, or accumulate upstream events into a single JSON response. A client bridge may also poll task results and construct a downstream stream from newly generated text. The downstream response format therefore need not match how the upstream product delivers its answer. Applications can consume the resulting inference through their expected protocol while the project handles product-specific execution and output retrieval.

\noindent\textbf{Resource management.}
Resource management (\ding{178}) determines which upstream resources serve a request and how execution proceeds when they become unavailable. Account selection may consider model eligibility, recorded quota, and account status; concurrency controls regulate simultaneous requests or client instances. Recovery actions depend on the failure: an expired credential may require renewal, invalid session state may require reconstruction, and temporary unavailability may lead to queuing, retrying, or returning an error. When an allowance is exhausted, a project may propagate the quota error or select another eligible account, either for a retry or for subsequent requests. With multiple eligible free accounts configured, one downstream API can draw on their separate allowances. 
This scheduling extends the capacity available through that API beyond a single account, subject to the eligibility and limits enforced by the upstream service.

\begin{table}[htbp]
\centering
\caption{Mechanisms recorded across 1,588 inference paths, grouped by upstream invocation method.}
\label{tab:lsf-mechanism-counts}
\begingroup
\newcommand{\lsfheatcell}[3]{\cellcolor[HTML]{#1}\textcolor{#2}{#3}}
\resizebox{1\linewidth}{!}{
\begin{tabular}{lccc}
\hline
\textbf{Mechanism}
& \textbf{Direct}
& \textbf{Browser}
& \textbf{Client} \\
\hline

\rowcolor{blue!6}
\multicolumn{4}{l}{\textbf{1. Downstream API handling}} \\
Authentication \& access control
& \lsfheatcell{DA141E}{white}{480}
& \lsfheatcell{FFF0A7}{black}{67}
& \lsfheatcell{FFFAC0}{black}{22} \\
Request validation
& \lsfheatcell{FFFCC5}{black}{14}
& \lsfheatcell{FFFFCC}{black}{2}
& \lsfheatcell{FFFFCC}{black}{1} \\
\hline

\rowcolor{blue!6}
\multicolumn{4}{l}{\textbf{2. Upstream identity and access}} \\
Upstream login \& authorization
& \lsfheatcell{FFEC9F}{black}{81}
& \lsfheatcell{FFF8BB}{black}{31}
& \lsfheatcell{FFFFFF}{black}{0} \\
Credential import \& persistence
& \lsfheatcell{FEB44E}{black}{227}
& \lsfheatcell{FFF0A7}{black}{67}
& \lsfheatcell{FFFECB}{black}{4} \\
Credential refresh \& renewal
& \lsfheatcell{FEE289}{black}{119}
& \lsfheatcell{FFFEC9}{black}{5}
& \lsfheatcell{FFFFCC}{black}{1} \\
Client identity \& header adaptation
& \lsfheatcell{FD9941}{black}{283}
& \lsfheatcell{FFFECB}{black}{4}
& \lsfheatcell{FFFFFF}{black}{0} \\
Access challenge handling
& \lsfheatcell{FEC965}{black}{186}
& \lsfheatcell{FFFDC8}{black}{9}
& \lsfheatcell{FFFFFF}{black}{0} \\
\hline

\rowcolor{blue!6}
\multicolumn{4}{l}{\textbf{3. Session and workflow management}} \\
Upstream session creation \& cleanup
& \lsfheatcell{FEB44E}{black}{227}
& \lsfheatcell{FFFDC6}{black}{11}
& \lsfheatcell{FFFECB}{black}{4} \\
Session binding \& reuse
& \lsfheatcell{FFE590}{black}{108}
& \lsfheatcell{FFF8BA}{black}{33}
& \lsfheatcell{FFFFCC}{black}{2} \\
Product workflow orchestration
& \lsfheatcell{FEDC7C}{black}{144}
& \lsfheatcell{FFFAC0}{black}{23}
& \lsfheatcell{FFFFFF}{black}{0} \\
\hline

\rowcolor{blue!6}
\multicolumn{4}{l}{\textbf{4. Request adaptation}} \\
Model mapping \& adapter selection
& \lsfheatcell{FEDC7C}{black}{142}
& \lsfheatcell{FFF9BE}{black}{25}
& \lsfheatcell{FFFECB}{black}{3} \\
Request field \& message conversion
& \lsfheatcell{EA2920}{black}{440}
& \lsfheatcell{FFFAC0}{black}{24}
& \lsfheatcell{FFFECB}{black}{4} \\
Prompt \& conversation reconstruction
& \lsfheatcell{800026}{white}{616}
& \lsfheatcell{FED977}{black}{153}
& \lsfheatcell{FFF9BE}{black}{25} \\
Tool-call emulation
& \lsfheatcell{FEDA78}{black}{151}
& \lsfheatcell{FFF3AF}{black}{52}
& \lsfheatcell{FFFEC9}{black}{7} \\
Attachment upload \& reference conversion
& \lsfheatcell{FFFDC8}{black}{9}
& \lsfheatcell{FFFDC6}{black}{10}
& \lsfheatcell{FFFFFF}{black}{0} \\
Product-specific protocol encoding
& \lsfheatcell{FFE590}{black}{107}
& \lsfheatcell{FFFFCC}{black}{2}
& \lsfheatcell{FFFFFF}{black}{0} \\
\hline

\rowcolor{blue!6}
\multicolumn{4}{l}{\textbf{5. Upstream invocation}} \\
Direct product request
& \lsfheatcell{800026}{white}{1322}
& \lsfheatcell{FFFFFF}{black}{0}
& \lsfheatcell{FFFFFF}{black}{0} \\
Browser-mediated request
& \lsfheatcell{FFFFFF}{black}{0}
& \lsfheatcell{FFE590}{black}{228}
& \lsfheatcell{FFFFFF}{black}{0} \\
Dedicated-client bridge
& \lsfheatcell{FFFFFF}{black}{0}
& \lsfheatcell{FFFFFF}{black}{0}
& \lsfheatcell{FFFBC2}{black}{38} \\
\hline

\rowcolor{blue!6}
\multicolumn{4}{l}{\textbf{6. Response adaptation}} \\
Upstream output parsing \& extraction
& \lsfheatcell{FFF6B6}{black}{39}
& \lsfheatcell{FFF9BD}{black}{28}
& \lsfheatcell{FFFFFF}{black}{0} \\
Response schema conversion
& \lsfheatcell{FC552C}{black}{376}
& \lsfheatcell{FFEDA0}{black}{77}
& \lsfheatcell{FFFCC4}{black}{15} \\
Streaming protocol conversion
& \lsfheatcell{EF3323}{black}{425}
& \lsfheatcell{FFF1AB}{black}{59}
& \lsfheatcell{FFFEC9}{black}{5} \\
Response buffering \& stream aggregation
& \lsfheatcell{FC5D2E}{black}{367}
& \lsfheatcell{FFF1AB}{black}{58}
& \lsfheatcell{FFFEC9}{black}{7} \\
Error \& termination status mapping
& \lsfheatcell{FED16E}{black}{169}
& \lsfheatcell{FFF6B6}{black}{40}
& \lsfheatcell{FFFEC9}{black}{7} \\
\hline

\rowcolor{blue!6}
\multicolumn{4}{l}{\textbf{7. Resource management}} \\
Account selection \& rotation
& \lsfheatcell{FD9941}{black}{283}
& \lsfheatcell{FFFBC2}{black}{18}
& \lsfheatcell{FFFECB}{black}{3} \\
Quota checks \& account availability
& \lsfheatcell{FFFAC0}{black}{22}
& \lsfheatcell{FFFFCC}{black}{1}
& \lsfheatcell{FFFFCC}{black}{2} \\
Concurrency \& rate control
& \lsfheatcell{FFF3AE}{black}{53}
& \lsfheatcell{FFF0A8}{black}{63}
& \lsfheatcell{FFFECB}{black}{3} \\
Request retry \& failover
& \lsfheatcell{FD8C3C}{black}{310}
& \lsfheatcell{FFF5B5}{black}{41}
& \lsfheatcell{FFFEC9}{black}{6} \\
Usage estimation \& accounting
& \lsfheatcell{FFF1A9}{black}{62}
& \lsfheatcell{FFFEC9}{black}{6}
& \lsfheatcell{FFFFFF}{black}{0} \\
Inference result caching
& \lsfheatcell{FFFECB}{black}{3}
& \lsfheatcell{FFFFCC}{black}{1}
& \lsfheatcell{FFFFFF}{black}{0} \\
\hline
\end{tabular}}
\par\smallskip
\begin{minipage}{\linewidth}
\footnotesize
\textit{Note.} Columns represent direct product requests, browser-mediated requests, and dedicated-client bridges, respectively.
\end{minipage}
\endgroup
\end{table}

\subsubsection{Mechanism distribution}
\autoref{tab:lsf-mechanism-counts} summarizes the 30 mechanisms identified across 1,588 inference paths, organized into seven functional groups and separated by upstream invocation method. 
We found that API compatibility depends on transformations implemented by \serv{} projects. Prompt and conversation history reconstruction was the most common adaptation mechanism, recorded on 794 of 1,588 paths (50.0\%) (\autoref{tab:lsf-mechanism-counts}). Tool-call emulation appeared on 210 paths. In \href{https://github.com/0x775/llm-bridge/blob/b0948e5b155f609b99923fb80defc047f98e98eb/main.go\#L446}{\texttt{0x775/llm-bridge}}, generated tool tags are parsed into the downstream \texttt{tool\_calls} field. Buffering or aggregation, recorded on 432 paths, changes how output is delivered. The non-streaming path in \href{https://github.com/jonnyquan/deepseek-free-api/blob/737668eecbf31a2b36d3efd1e4fb4dc6c47ecd87/src/api/controllers/chat.ts\#L596}{\texttt{jonnyquan/deepseek-free-api}} accumulates upstream SSE content into a complete JSON response. These implementations show how \serv{} projects supply downstream protocol features through their own handling of conversation history, tool-call representations, and generated output.

Upstream access also relies on reusable authentication state. We recorded credential import and persistence on 298 paths, compared with 112 for login and authorization. In \href{https://github.com/Alastor418/deepseek-2api-gui/blob/ef682ab553204b61641e4d473788a591124d10b8/browser-extensions/chrome/background.js\#L34}{\texttt{Alastor418/deepseek-2api-gui}}, a browser extension transfers authorization values and cookies into an account pool for subsequent inference requests. Account selection and rotation appeared on 304 paths, and retry and failover on 357. Their interaction determines which allowance is available during recovery. In \href{https://github.com/jonnyquan/deepseek-free-api/blob/737668eecbf31a2b36d3efd1e4fb4dc6c47ecd87/src/api/controllers/chat.ts\#L596}{\texttt{jonnyquan/deepseek-free-api}}, the project randomly selects an initial credential and retains it across retries. With multiple credentials configured, separate requests can use different account allowances, while retries of an individual request remain tied to the initial account. Stytch reported a related case at Kilo Code, where attackers created thousands of accounts to claim free credits, then resold those accounts or routed LLM traffic through them for profit~\cite{stytch_kilocode_abuse}. Stripe likewise reported suspected multi-account abuse in 7.4\% of customer sign-ups at AI companies in its analysis~\cite{stripe_fraud_2026}. 

\subsection{Case Study}

We examine two \serv{} projects to illustrate how the mechanisms identified in \autoref{sec:mechanism} expose free product inference through independent APIs. We traced one inference path at a fixed revision of each project.  \href{https://github.com/aurora-develop/aurora/tree/cbfbabc9c14c7278e4dd91cf8807b2bf500860b1}{\texttt{aurora-develop/aurora}} submits requests directly to ChatGPT Web, while \href{https://github.com/lbjlaq/Antigravity-Tools-LS/tree/d312237af83820ca15a636e81c5e61660bdf13f4}{\texttt{lbjlaq/Antigravity-Tools-LS}}  invokes Antigravity through its local language server, covering two upstream invocation methods identified in \autoref{fig:lsf_sankey}.

\begin{figure}[t]
    \centering
    \includegraphics[width=1\linewidth]{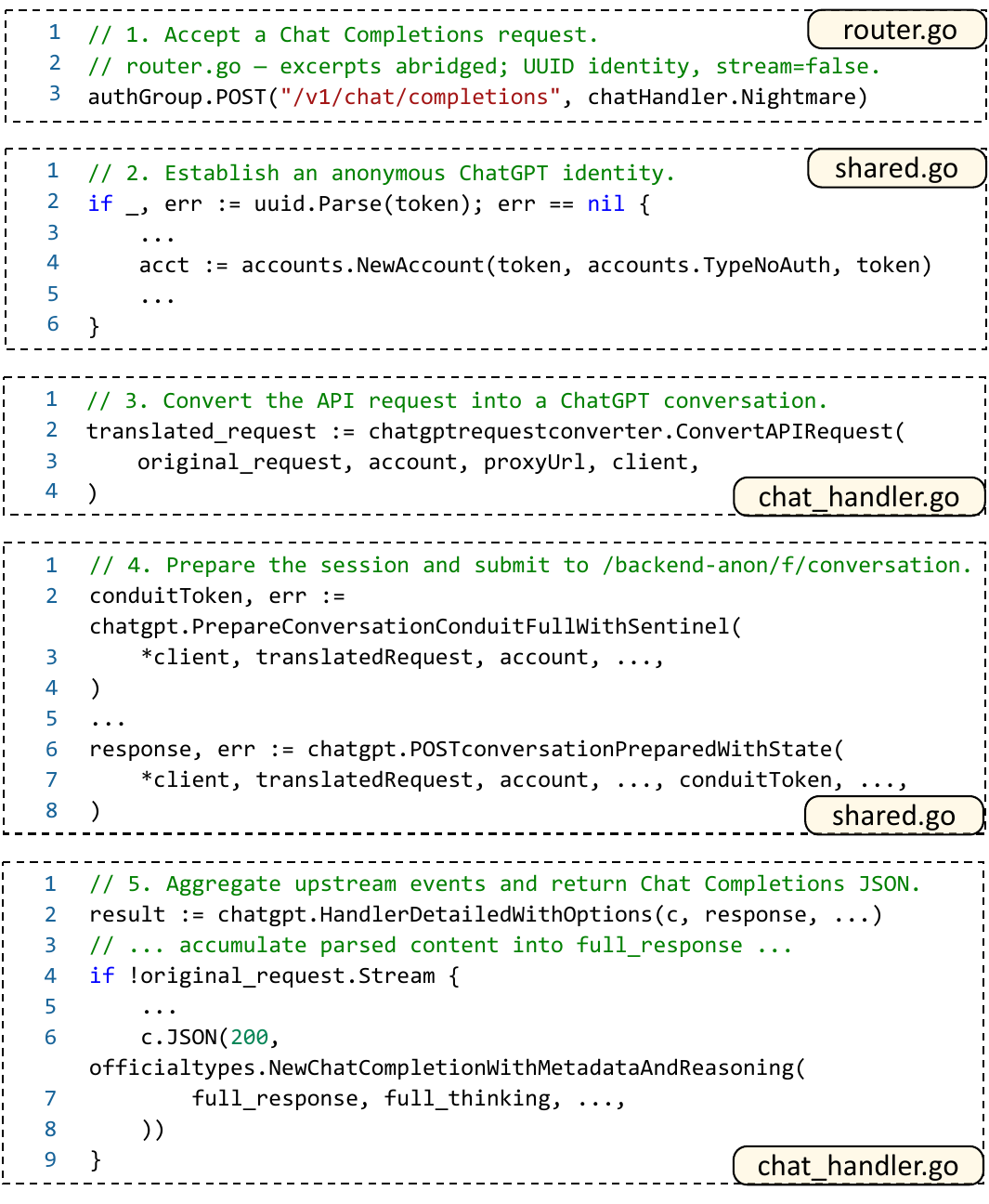}
    \caption{Aurora repackages anonymous ChatGPT inference as an API.}
    \label{fig:case_aurora}
\end{figure}

\noindent\textbf{Anonymous product access.} The examined path in \href{https://github.com/aurora-develop/aurora/tree/cbfbabc9c14c7278e4dd91cf8807b2bf500860b1}{\texttt{aurora-develop/aurora}} repackages ChatGPT Web's anonymous free access as an OpenAI-compatible Chat Completions API. As shown in \autoref{fig:case_aurora}, a non-streaming request to \texttt{/v1/chat/completions} supplies a UUID that the implementation maps to an anonymous product identity. Aurora converts the messages into a web conversation request, performs the product's session preparation, and submits the request directly to the anonymous conversation endpoint. Although the downstream request specifies a single JSON response, ChatGPT Web returns conversation events incrementally. Aurora parses these events, accumulates the generated text, and returns the assembled answer as a Chat Completions JSON response over the original connection.
This path connects session handling and response adaptation: an anonymous product session supplies free inference, while the project reconstructs the answer in the downstream protocol. Applications can therefore consume this inference through a standard API without credentials for a registered ChatGPT account. The relevant restriction concerns output extraction: the OpenAI consumer terms reviewed in \autoref{tab:provider-policy-restrictions} prohibit programmatic extraction of generated output, which Aurora performs when parsing the web response to construct the downstream answer.

\begin{figure}[htbp]
    \centering
    \includegraphics[width=1\linewidth]{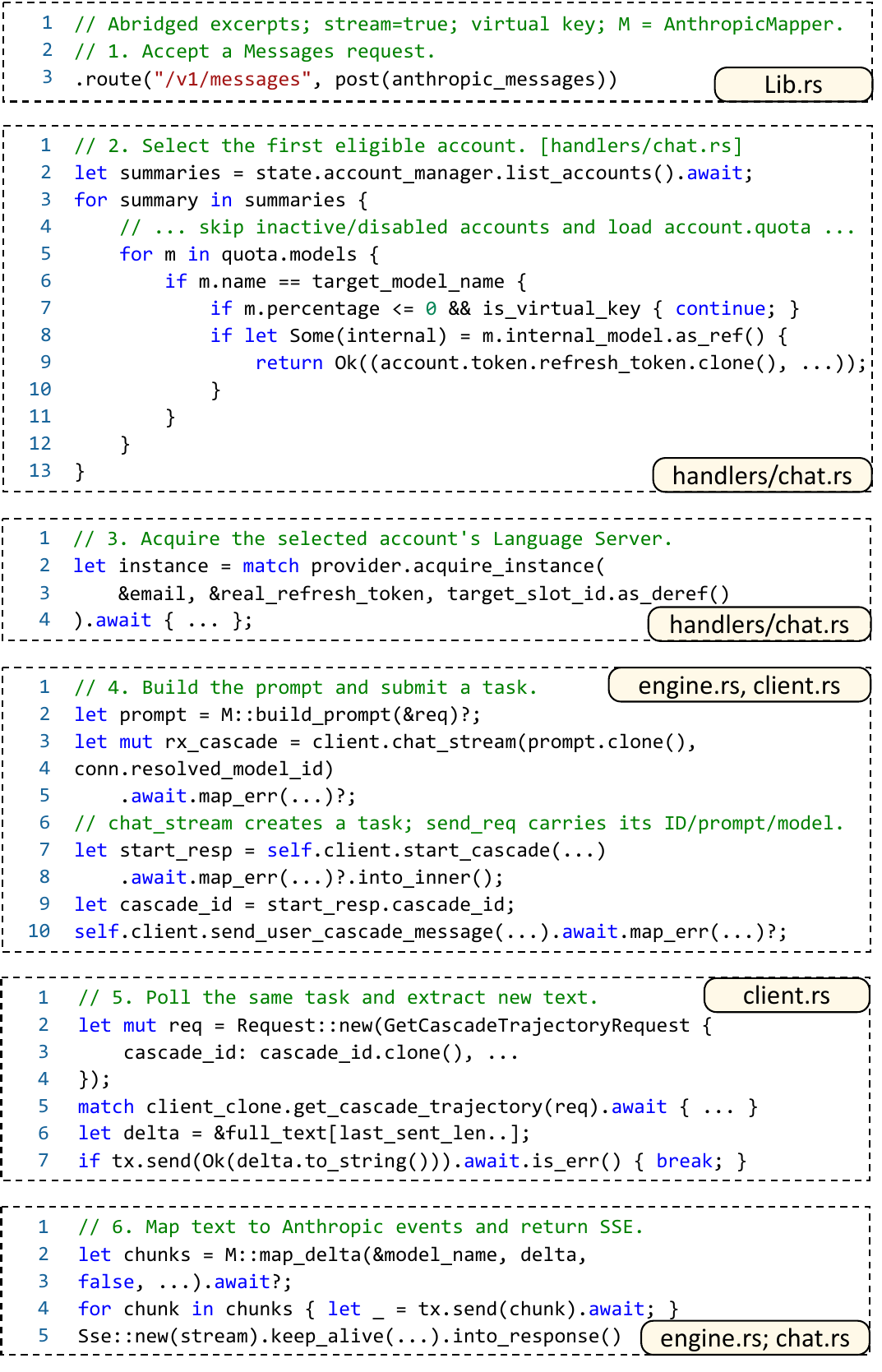}
    \caption{Antigravity-Tools-LS repackages free inference through Antigravity's Language Server.}
    \label{fig:case_lbjlap}
\end{figure}

\noindent\textbf{Account pooling and client bridging.} The examined configuration of \href{https://github.com/lbjlaq/Antigravity-Tools-LS/tree/d312237af83820ca15a636e81c5e61660bdf13f4}{\texttt{lbjlaq/Antigravity-Tools-LS}} exposes inference available to eligible free Antigravity accounts through an Anthropic-compatible Messages API. As shown in \autoref{fig:case_lbjlap}, requests authenticated with a virtual API key issued by the project are routed through a pool of managed product accounts. The selector skips inactive accounts and those with proxy access disabled, then selects the first eligible account with positive recorded quota for the requested model. Its credentials are used to obtain a language server instance. The project reconstructs the prompt from the incoming messages and submits it with the selected model through the product's local Cascade gRPC interface. It then polls the trajectory associated with the same Cascade identifier, extracts newly generated text, and converts these increments into Anthropic events returned over the original HTTP connection.
This implementation combines resource scheduling with the dedicated-client bridge identified in \autoref{fig:lsf_sankey}. Multiple eligible free accounts can supply inference through one downstream API. When an account's recorded model quota is exhausted, subsequent requests can be assigned to another eligible account. The language server handles product execution, while the project constructs the downstream stream from successive task results. In \autoref{tab:provider-policy-restrictions}, Antigravity terms prohibit access through third-party software. Where these terms apply, and no separate authorization exists, driving the language server through an independent API conflicts with that restriction.

\section{Discussion}
\label{sec:discussion}

\noindent\textbf{Security implications.}
Our findings inform both providers' protection of free inference resources and users' assessment of LSF deployments. Providers should explicitly address automated product access, aggregation of free allowances, and redistribution through third-party interfaces in their service policies. The observed invocation and account-management mechanisms provide concrete scenarios for evaluating entitlement checks across direct requests, browser-mediated access, and dedicated-client bridges, including whether account pooling circumvents restrictions on individual beneficiaries. The identified repositories and upstream channels also provide leads for investigating downstream abuse. For users, third-party-operated services introduce an operator that processes their prompts and responses, while account holders supplying credentials additionally delegate upstream access. Self-hosting reduces reliance on an external operator but still requires trust in the implementation's credential handling, logging, and conversation isolation. Use that violates provider conditions may also result in account restrictions, suspension, or termination. We therefore recommend that users avoid deploying or relying on LSF services and obtain inference through provider-authorized interfaces and access arrangements.

\noindent\textbf{Responsible disclosure.}
We observed the following GitHub warning on 2 repository pages when we visited them while signed in:\footnote{When we visited the same pages while signed out, GitHub redirected us to its sign-in page. GitHub's \href{https://docs.github.com/en/site-policy/acceptable-use-policies/github-active-malware-or-exploits}{\emph{Active Malware or Exploits}} policy states that, in rare cases of widespread abuse of dual-use content, GitHub may restrict access to the content, typically by requiring users to sign in.} \textit{``This repository contains malicious content that may cause technical harm. We have decided to preserve this content for security research purposes. Please exercise CAUTION when clicking links, downloading releases, or otherwise interacting with this repository.''} These moderation actions indicate that some \serv{} projects have already drawn scrutiny from GitHub.
We reported all 513 identified \serv{} projects to GitHub for review. We also notified the providers of the 64 upstream LLM services, providing the relevant repository links, source-code locations documenting the inference paths, and descriptions of the mechanisms used to expose free inference through independent APIs. These reports highlighted potential conflicts with the providers' restrictions on automated access, account use, and service redistribution. We also recommended validating free-access eligibility at inference endpoints, limiting credential reuse, and monitoring coordinated consumption of free allowances across accounts.

\noindent\textbf{Threats to validity.}
Our discovery process depends on search queries, publicly available repositories, and the downstream protocol families selected for analysis. It may miss projects with unfamiliar terminology, custom interfaces, inaccessible source code, or implementations outside GitHub. Manual validation of retrieved repositories cannot recover projects that were never discovered. The dataset therefore characterizes the observed open-source ecosystem without establishing its complete size or the prevalence of LSF across all third-party inference services. Finally, free-access offerings, provider policies, and project implementations may evolve over time, so our findings only reflect the current state during the study.

\section{Related Work}
\label{sec:related}

\noindent\textbf{Security of LLM proxy services.}
Prior studies have examined the trustworthiness of third-party LLM services, including model identities, response consistency, and billing transparency~\cite{zhang2026shadowapis,lin2026gatescope,wu2026middleman}. Related auditing methods identify the models behind inference interfaces and detect deviations from reference model outputs~\cite{pasquini2025llmmap,gao2025modelequality}. Other research investigates incentives for overcharging under token-based pricing and methods for verifying hidden reasoning-token usage~\cite{velasco2026overcharging,sun2026coin}. Studies of malicious intermediaries further examine response manipulation, credential exposure, and the misuse of upstream resources~\cite{liu2026youragent}, while CacheTracer uses prefix-cache side channels to infer hidden dependencies among API resellers~\cite{ji2026cachetracer}. Our work complements these studies by investigating the entitlement-circumvention mechanisms that support freeloading services.

\noindent\textbf{Misuse of LLM applications.}
Prior work has investigated how LLM capabilities are manipulated or packaged for harmful and unintended uses. Studies of jailbreaks examine prompts circulating in online communities and automated methods for inducing prohibited outputs~\cite{shen2024dan,zhou2025jailbreak,deng2024masterkey}. Measurements of malicious LLM services and misused GPTs characterize their deployment, safeguard evasion, and effectiveness in supporting harmful activities~\cite{lin2024malla,shen2025gptracker}. Broader studies of LLM application ecosystems investigate harmful behavior, privacy violations, and app impersonation and cloning~\cite{hou2025insecurity,xie2025squatting,zhao2025llmapp}. Research on application capability boundaries also shows that specialized LLM applications can be induced to perform tasks outside their intended functions, even when those tasks do not violate the underlying model's safety constraints~\cite{zhang2026beyondjailbreak}. We examine a complementary dimension of LLM misuse, focusing on how actors circumvent access and resource constraints to obtain inference for themselves or downstream users.

\noindent\textbf{Abuse of inference resources.}
Resource-oriented attacks exploit the computational and financial costs of inference. Adversarial inputs can prolong generation or amplify reasoning, increasing resource consumption and response latency~\cite{dong2025engorgio,li2026thinktrap,kumar2025overthink}. Research on LLM-based agents further examines resource-exhaustion vulnerabilities~\cite{luo2026agentdos} and the diversion of users' computational budgets to attacker-selected workloads~\cite{zhang2025leechhijack}. Complementary measurement studies characterize publicly exposed LLM deployments, access-control weaknesses, and exploitable service interfaces~\cite{hou2026mapping,liu2026lens}, while honeypot-based research documents attack attempts against exposed inference infrastructure~\cite{elzer2026ollamadrama}. While prior work primarily focuses on attacks against inference services, our study investigates the freeloading mechanisms through which abusive services circumvent upstream usage constraints to abuse inference resources.
\section{Conclusion}
\label{sec:conclusion}

We presented a systematic measurement study of LLM service freeloading in the wild, analyzing 513 open-source projects across 64 upstream LLM services. Our analysis characterized the projects' growth, code reuse, and development activity, and identified how they combine credential management, request and response adaptation, and account scheduling to expose free product inference through independent APIs. Besides, our policy review found that 58 of the 62 services with available terms explicitly prohibited at least one practice associated with LSF. Our study advances the understanding of LSF and informs efforts to protect free inference resources and their users.

\section*{Ethics Considerations}

Our study analyzes public repositories and provider documentation to improve protections against LLM service freeloading. Our data-handling protocol limits collection to relevant technical information and excludes credentials and unrelated personal data from released materials. We distinguish implementation capabilities from policy violations and avoid attributing malicious intent to developers. Our findings and tools could facilitate unauthorized resource use; release review therefore addresses sensitive operational details, accompanying mitigations, and remediation timing. As of paper submission, we have disclosed the identified issues to the affected LLM providers. These measures balance the benefits of identifying weaknesses in service controls against risks to providers, developers, and users.

\section*{Use of Generative AI}

We use ChatGPT for language and presentation editing, and Codex for implementation, refactoring, and debugging. The authors review AI-assisted text and inspect AI-assisted code. Generative AI also assists repository inspection; classifications and mechanism descriptions are verified against source code and service documentation. The authors verify references, results, and conclusions and take full responsibility for the work’s accuracy, originality, and integrity.


\bibliographystyle{IEEEtranS}
\bibliography{main}

\appendices

\end{document}